\documentclass[journal=mamobx,manuscript=article,layout=traditional]{achemso}
\setkeys{acs}{articletitle=true}
\usepackage[version=3]{mhchem} 
\usepackage[T1]{fontenc}       

\usepackage{newtxmath}
\usepackage{newtxtext}
\usepackage{bm} 
\usepackage[labelfont=bf]{caption} 

\usepackage[hyperfootnotes=false]{hyperref}	 
\hypersetup{
  colorlinks,
  citecolor=blue,
  linkcolor=blue,
  urlcolor=blue
}

\usepackage{titlesec} 
\titleformat{\subsection}[runin]
{\normalfont\bfseries}{\thesubsection{.}}{1em}{}[.]

\titleformat{\section}
{\normalfont\sffamily\large\bfseries}{\thesection{.}}{1em}{\MakeUppercase}

\usepackage{calligra}
\usepackage{lineno}
\DeclareMathAlphabet{\mathcalligra}{T1}{calligra}{m}{n}

\usepackage{amsmath}
\usepackage{amssymb}
\usepackage{amsfonts}

\usepackage{graphicx}

\usepackage{array}
\usepackage{multirow}
\usepackage{color}
\usepackage{transparent}
\usepackage{float}
\usepackage[mathscr]{eucal}

\usepackage{natbib}

\newcommand{\kB}{k_\mathrm{B}}
\newcommand{\lB}{l_\mathrm{B}}
\newcommand{\ie}{{\it{i.e.}}}

\newcolumntype{P}[1]{>{\centering\arraybackslash}p{#1}}

\usepackage{soul}  
\SectionNumbersOn
\let\oldmaketitle\maketitle
\let\maketitle\relax

\title{\flushleft Charged dendrimers revisited: Effective charge and surface potential of dendritic polyglycerol sulfate}

\author{\rm\flushleft Xiao Xu}
\affiliation{\rm\small Institut f\"ur Weiche Materie und Funktionale Materialien, Helmholtz-Zentrum Berlin, Hahn-Meitner-Platz 1, 14109 Berlin, Germany}
\alsoaffiliation{\rm\small Institut f{\"u}r Physik, Humboldt-Universit{\"a}t zu Berlin, Newtonstr.~15, 12489 Berlin, Germany}
\alsoaffiliation{\rm\small Multifunctional Biomaterials for Medicine, Helmholtz Virtual Institute, Kantstr. 55, 14513 Teltow-Seehof, Germany}
\author{Qidi Ran}
\affiliation{\rm\small Institut f\"ur Weiche Materie und Funktionale Materialien, Helmholtz-Zentrum Berlin, Hahn-Meitner-Platz 1, 14109 Berlin, Germany}
\alsoaffiliation{\rm\small Multifunctional Biomaterials for Medicine, Helmholtz Virtual Institute, Kantstr. 55, 14513 Teltow-Seehof, Germany}
\alsoaffiliation{\rm\small Institut f\"ur Chemie und Biochemie, Freie Universit\"at Berlin, Takustr. 3, 14195 Berlin, Germany}
\author{Rainer Haag}
\affiliation{\rm\small Multifunctional Biomaterials for Medicine, Helmholtz Virtual Institute, Kantstr. 55, 14513 Teltow-Seehof, Germany}
\alsoaffiliation{\rm\small Institut f\"ur Chemie und Biochemie, Freie Universit\"at Berlin, Takustr. 3, 14195 Berlin, Germany}
\author{Matthias Ballauff}
\affiliation{\rm\small Institut f\"ur Weiche Materie und Funktionale Materialien, Helmholtz-Zentrum Berlin, Hahn-Meitner-Platz 1, 14109 Berlin, Germany}
\alsoaffiliation{\rm\small Institut f{\"u}r Physik, Humboldt-Universit{\"a}t zu Berlin, Newtonstr.~15, 12489 Berlin, Germany}
\alsoaffiliation{\rm\small Multifunctional Biomaterials for Medicine, Helmholtz Virtual Institute, Kantstr. 55, 14513 Teltow-Seehof, Germany}
\author{Joachim Dzubiella}
\affiliation{\rm\small Institut f\"ur Weiche Materie und Funktionale Materialien, Helmholtz-Zentrum Berlin, Hahn-Meitner-Platz 1, 14109 Berlin, Germany}
\alsoaffiliation{\rm\small Institut f{\"u}r Physik, Humboldt-Universit{\"a}t zu Berlin, Newtonstr.~15, 12489 Berlin, Germany}
\alsoaffiliation{\rm\small Multifunctional Biomaterials for Medicine, Helmholtz Virtual Institute, Kantstr. 55, 14513 Teltow-Seehof, Germany}
\email{joachim.dzubiella@helmholtz-berlin.de}

\begin{document}

\pagenumbering{arabic}
\noindent

\parindent=0cm
\setlength\arraycolsep{2pt}


\oldmaketitle

\begin{abstract}
We investigate key electrostatic features of charged dendrimers at hand of the biomedically important dendritic polyglycerol sulfate (dPGS) macromolecule using multi-scale computer simulations and Zetasizer experiments. In our simulation study, we first develop an effective mesoscale Hamiltonian specific to dPGS based on input from all-atom, explicit-water  simulations of dPGS of low generation. Employing this in coarse-grained, implicit-solvent/explicit-salt Langevin dynamics simulations, we then study dPGS structural and electrostatic properties up to the sixth generation.  By systematically mapping then the calculated electrostatic potential onto the Debye-H{\"u}ckel form -- that serves as a basic defining equation for the effective charge -- we determine well-defined effective net charges and corresponding radii, surface charge densities, and surface potentials of dPGS. The latter are found to be up to one order of magnitude smaller than the  bare values and consistent with previously derived theories on charge renormalization and weak saturation for high dendrimer generations (charges).  Finally, we find that the surface potential of the dendrimers estimated from the simulations compare very well with our new electrophoretic experiments. 

\vspace{5ex}
\end{abstract}

\maketitle
\setlength\arraycolsep{2pt}
\begin{tocentry}
\includegraphics[scale=0.23, bb=-60 0 948 445]{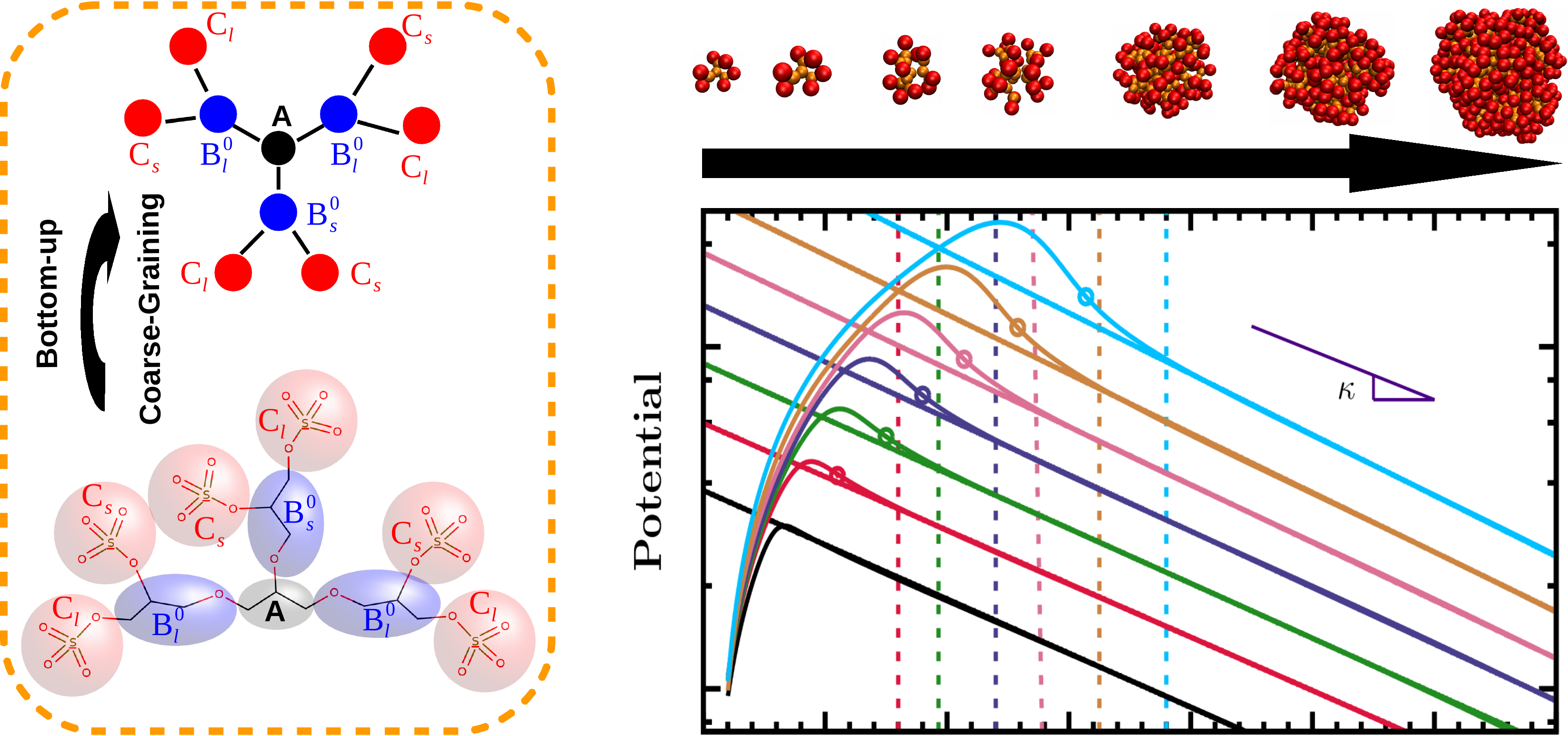}
\end{tocentry}

\section{Introduction}

Charged dendritic macromolecules have attracted strong and broad interest from both academic and industrial researchers due to their versatile bioapplications, such as drug delivery, tissue engineering, and biological imaging~\cite{Ballauff2004,  Lee:review, Tian2013}.
Recently, a high potential candidate for the use in future medical treatments has been identified based 
on dendritic polyglycerol sulfate (dPGS).~\cite{Khandare2012}  The latter has been found very efficient for the treatment of neurological disorders arising from inflammation~\cite{Maysinger2015}, therapeutics for preventing the process of tissue damage~\cite{Reimann2015}, acting as delivery platform~\cite{Groeger2013, Vonnemann2014}, e.g., transporting to tumor cells,~\cite{Sousa-Herves2015} and as imaging agent for the diagnosis of rheumatoid arthritis~\cite{Vonnemann2014}.   Due to its charged terminal groups dPGS interacts mainly through electrostatics. The high anionic surface charge is therefore basis for dPGS' high anti-inflammatory potential.~\cite{JensDernedde2010,Hoshino2014}.

The important applications of dendrimeric macromolecules have initiated large efforts in their detailed  microscopic characterization by theory and computer simulations~\cite{Ballauff2004, Tian2013}.   While there is no simulation work yet in literature characterizing the basic structural features of dPGS,  a large number of atomistic computer simulations, for example, of PAMAM [poly(amidoamine)]-based dendrimers have been  performed.~\cite{Maiti2004,Naylor1989, Lee2002, Han2005, Maiti2005, Maiti2008}   On the other hand, to overcome the limitation of the system size of atomistic simulations, coarse-grained (CG) monomer-resolved models with more or less inclusion of specific chemical features have led to plentiful structural insight~\cite{Muratt1996,Welch1998, Lyulin2004:1, Lyulin2004, Giupponi2004, Lee2006,Lee2008,Lee2009, Chong2015, Gurtovenko2006, Blaak:2008, Carbone2010, Huismann2010, Huismann2010B, Tian2012,  Das2014, Welch2000, Lenz2012,  Klos2009, Maiti2009, Tian2011, Klos2010, Klos2011, Klos2013,  Huismann2012, Tian2010}.  For the case of (internally and surface) charged dendrimers, one focus has been set on the dominant role of condensed counterions and charge renormalization~\cite{Ohshima1982, Alexander1983, Ramanath1988, Belloni1998, Bocquet2002, Netz2003,Manning2007,DavidA.J.Gillespie2014} in modulating the conformation and effective charge of the dendrimers.~\cite{Klos2010, Klos2011, Klos2013, Huismann2012, Tian2010} 

However, it has been hardly attempted to consistently calculate the effective surface potential (and its location) of charged dendrimers so far, despite its significance for electrostatic interactions. One reason could be that the identification of condensed counterions requires the definition of a cut-off region in space that contains condensed-types of ions distinct from those in the diffusive double layer, with highly varying definitions in the just cited literature.  As a consequence, effective charges and the spatial delimitation of diffusive double-layer behavior have been inconsistently defined, hampering a meaningful comparison to analytical theory and experiments.  Regarding specifically the calculation of surface potentials, one notable exception is the simulation report on charged PAMAM dendrimers that revealed a superlinear increase of the effective charge and the surface potential with generation number~\cite{Maiti2008}. In view of the theoretically predicted saturation (or at least very weak, sublinear increase) of the effective charge with bare charge for simple charged spheres~\cite{Ohshima1982, Alexander1983, Ramanath1988, Belloni1998, Bocquet2002, Netz2003,Manning2007, DavidA.J.Gillespie2014}, this result, however, is unexpected and not well understood. In fact, in the case of carboxyl-terminated dendrimers capillary electrophoretic experiments demonstrated that higher generation dendrimers (generation 5) even have a smaller effective surface charge than lower generations (generation 2).~\cite{Huang2000} Hence, despite the large body of studies on charged dendrimers in the last two decades, key electrostatic features have not been yet consistently addressed. 

Driven by the urgent need to develop accurate modeling tools and interpretation for the interactions between charged dendritic drugs that are predominantly of electrostatic nature, we here investigate dPGS with a particular focus on the determination of its electrostatic surface properties.  Compared to the relatively large and steady growing number of experimental publications on dPGS, a deeper molecular characterization of dPGS is still lacking. Hence, in this paper, based on all-atom MD simulations, a CG model specific to dPGS  is reported for its future modeling in biological environments (e.g., interacting with proteins or membranes).  In particular, we introduce a simple but accurate scheme how to systematically calculate well-defined effective surface charges and potentials of charged dendrimers in the case of dPGS based on the most practical definition by mapping  the calculated potentials directly to the Debye-H{\"u}ckel potential in the far-field regime.  We compare them to available theories and new experimental $\zeta$-potential measurements (also included in this contribution) with consistent outcome. Our study thus paves the way for future simulations and interpretations of the dPGS' and related dendritic polyelectrolytes' action in biological context (e.g., interacting with proteins or membranes) to understand and optimize their proven selective binding properties and efficacy in the medical treatment of inflammatory diseases.




\section{Models and methods}

\subsection{All-atom model and explicit-water MD simulations}

\begin{figure*}[ht]
\includegraphics[scale=0.5]{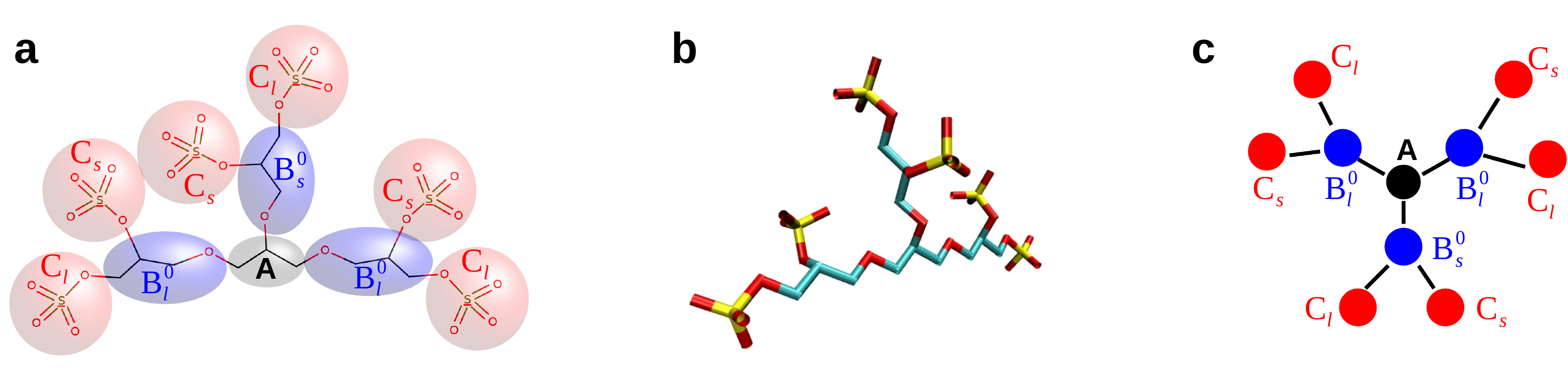}
\caption{
(a) Chemical formula, (b) atomistic structure, and (c) coarse-grained (CG) structure sketch of the zeroth generation G$_0$--dPGS molecule. 
In panel (a), we depict the atomistic subunits C$_3$H$_5$, C$_3$H$_7$O, and SO$_4^-$  corresponding 
to the CG bead types  A, B$_i$, and C$_i$ ($i=s,l$) in panel (c) by the black, blue, red colored regions, respectively. 
}
\label{ff}
\end{figure*}

We start the development of the dPGS coarse-grained (CG) force--field from fully atomistic (explicit-water) MD simulations of dPGS of 
generations $n=0$ and 1 (i.e., G$_0$ and G$_1$) from which effective intra-bead potentials are derived.  The constituting elements are the initiator core C$_3$H$_5$,  repeating side chain units C$_3$H$_5$O, as well as the terminal sulfate groups, see the exemplary chemical structure of  G$_0$, in Fig.~1.  All our simulations are carried out by the {GROMACS 4.5.4} software package \cite{Hess2008},
where the {GROMOS 45a3} force field~\cite{Schuler2001} is applied. The latter is optimized to model lipids with long aliphatic chains or branches, which makes it an appropriate force field to model dPGS. The water is represented by the extended simple point charge (SPC/E) model~\cite{Berendsen1987}. 
The partial charges of dPGS are calculated according to the Gaussian 09 software~\cite{Frisch} with the cc-PVTZ DFT basis set and used here in combination with the {GROMOS 45a3} force field. The assigned partial charges among the glycerol groups stay close to the re-optimized parameters for ethers.~\cite{Horta2011} And those for the sulfate group agree with recent work~\cite{Jozic2013} which have been applied, for instance, in the simulation of sodium dodecyl sulfate micelles.~\cite{Tang2014} The calculations of partial charges of the dPGS atoms and the influence of their particular choices on the dendrimeric structure are summarized in the Supplementary Information (SI).

The initial  configurations of G$_0$ and G$_1$-dPGS are constructed in vacuum with the program ChemDraw~\cite{Evans2014}.
A single dendrimer is then placed in the center of a cubic simulation cell with an initial side length $L = 7.9$~nm with 
periodic boundary conditions in all three directions.  The dendrimer is subsequently hydrated with 16543 water molecules.   To preserve the overall charge neutrality of G$_0$ and G$_1$, six or twelve Na$^+$-counterions are added, respectively.  The electrostatic interactions are calculated via the Particle-Mesh-Ewald (PME)~\cite{Essmann1995} summation where the long-range potential is evaluated in the reciprocal space using the Fast-Fourier Transform (FFT) with a grid spacing of $0.16$~nm and a cubic interpolation of fourth order.  
A cut-off radius of 1~nm is defined for both PME summation and van der Waals real-space interactions. After a 100~ns equilibration in the isobaric $NPT$ ensemble at conditions of $P=1$~bar and $T=310$~K,  a production run of 1~$\mu$s generates a working trajectory in a canonical $NVT $ simulation. 
We utilize the Berendsen thermostat and the Rahman-Parrinello barostat.  
To integrate Newton's  equation of motion we employ the leap-frog algorithm with a time step of 2~fs.

\subsection{Coarse-graining procedure }

For our CG model we now define three coarse-grained bead types A, B, and C, chosen to be located at the center-of-mass position of repeating units, cf.~Fig.~1.  While type A  simply models the central core unit (C$_3$H$_5$),  the natural choice for B and C is reflected in the chemical 
formula C$_3$H$_5$(C$_3$H$_5$S$_2$O$_9$)$_3$ for G$_0$-dPGS which defines the repeating 
 units C$_3$H$_5$O as type B and the terminal group SO$_4$ as monomer C.     As one can see in the chemical structure in Fig.~1,
  the intra-bead potentials between B and C beads (as well as between B and B beads in higher generations) depend on how the corresponding atomistic groups are connected: in a given triplet   of units around a connecting central hub unit, two units feature an extra bond so that it is needed to introduce  `short' and `long'
 bead types B and C for intermediate branching cycles $m = 0..n$.  In the following we therefore distinguish between 
 B$_s^m$ and B$_l^m$ as well as C$_s$ and C$_l$ beads,  respectively. 
 Hence, for the coarse-grained force field we need to define bond potentials of types A-B$_j^{0}$, B$_i^m$-B$_j^{m+1}$, B$_i^n$-C$_j$, with $i=s,l$ and angle potentials for all relevant triplets, for example,  A-B$_i^0$-B$_j^1$, B$_i^{m}$-B$_j^{m+1}$-B$_k^{m+2}$, B$_i^{m+1}$-B$_j^{m}$-B$_k^{m+1}$, B$_i^{n-1}$-B$_j^{n}$-C$_k$, etc.  Note that there must be 'up-down' symmetry in the potentials, e.g.,  B$_i^m$-B$_j^{m+1}$=B$_i^m$-B$_j^{m-1}$ for the bond potentials and analogous rules for the angular potentials.  Finally, 
 non-bonded inter-bead interactions have to be defined between types A, B, and C. 
 
We employ harmonic potentials for the
intramolecular (bond and angular) interactions and use the Lennard-Jones (LJ) interaction for all inter-bead potentials. 
Additionally, the beads of type C and ions carry charges.  Hence, the CG force field can be formally summarized by the CG 
Hamiltonian
\begin{eqnarray}
U^{\rm CG} &=& \sum_{\rm bonds} \frac{1}{2}k_b(l -l_0)^2 + \sum_{\rm angle}\frac{1}{2} k_a (\theta - \theta_0)^2 \nonumber \\
 &+&
\sum_{i<j} 4\epsilon_{ij}\left[\left(\frac{\sigma_{ij}}{r_{ij}}\right)^{12}-\left(\frac{\sigma_{ij}}{r_{ij}}\right)^6\right] + U_{\rm elec} .
\label{H_intra}
\end{eqnarray}
where $r_{ij}$ is the bead--bead distance, $k_b$ and $k_a$ are the bond and angular spring constants, respectively, $l$ represents the distance between consecutive beads and $l_0$ is the equilibrium bond length. The variable $\theta$ refers to the angle formed by a triplet of consecutive beads and $\theta_0$ is the equilibrium value. 
Only the C bead, representing the terminal sulfate group, carries a bare Coulomb charge $q_s = -e$.
It follows that the net charge valency for the CG dPGS molecule of generation $n$ is $Z_{\rm bar} = -6(2^{n+1} - 2^{n})$ and thus the terminal beads number $N_{\rm ter} = |Z_{\rm bar}|$.
The electrostatic interactions for all charged beads (type C and ions) are included in $U_{\rm elec}$, via the Coulomb law
\begin{equation}
U_{\rm elec} = {\sum^{N_{\rm ter} + N_{\rm ion}}_{i=1} \sum^{N_{\rm ter} + N_{\rm ion}}_{j=1, j\ne i}}  {{l_B}\over {2r_{ij}}}.
\label{ele_CG}
\end{equation}
 The variable $\lB = e^2/$($4\pi \epsilon_0 \epsilon_r \kB T$) stands for the Bjerrum length, which is $\lB = 0.7$ nm in this study at body temperature $T= 310$ K and for water with a permittivity constant $\epsilon_r = 78.2$, $N_{\rm ion}$ denotes the number of ions, $e$ is the elementary charge, and $\epsilon_0$ is the permittivity of vacuum.   
 
We derive the bonded potentials by Boltzmann--inverting the corresponding target spatial distribution functions of the beads $f(x)$ we obtain from the atomistic MD simulation, via
\begin{equation}
U(x) = -\kB T\ln f(x), 
\label{Binver}
\end{equation}
where $x$ is either a bond length or angle variable and $f(x)$ is an equilibrium average over the fluctuations of all identical groups in the dendrimer. All potentials involving beads A and C are derived from simulations of G$_0$. All potentials involving beads only of type B are derived from simulations of G$_1$ as B-B bonds are absent in G$_0$.  The results and final parameters of the bonded CG potentials are discussed and  summarized in the next section. 

The extraction of the non-bonded A-A, B-B and C-C potentials directly from the atomistic MD simulation of G$_0$ or G$_1$-dPGS is very difficult due to the convoluted spatial structure of the dendritic dPGS. Therefore, a mapping using the  iterative Boltzmann inversion (IBI) scheme~\cite{Mueller-Plathe2002, Reith2003}  is out of reach.  
We therefore resort to the simplest approximation and perform explicit-water simulations of a one-component fluid of isolated A, B, and C monomers in explicit water, respectively, at relatively high dilution. The respective non-bonded pair potentials are then obtained by the simple Boltzmann--inversion according to Eq.~(3), where $f(x)$ then simply represents the radial distribution function. The charged sulfate monomers corresponding to subunit C are protonated to separate out approximately  the electrostatic monopole repulsion which later in the CG simulation are added again. 
The partial charges of the A, B, C chemical subunits are calculated according to the Gaussian 09 software~\cite{Frisch} with the cc-PVTZ DFT basis set, and used here in combination with the {GROMOS 45a3} force field, see the data in the SI.
We set the concentration of the one-component bead fluid to $c \sim 400$ mM  that is chosen high enough to obtain sufficient sampling and low enough to avoid large many-body effects and  possibly aggregated states. 
After obtaining the LJ parameters $\epsilon_{ii}$ and $\sigma_{ii}$ by fitting the LJ potentials to the obtained effective interaction (see next section) for the three  subunits, $i=$A,B,C, the corresponding values for the cross interactions are obtained by the conventional Lorentz-Berthelot mixing rules, i.e., $\sigma_{ij} = (\sigma_{ii}+\sigma_{jj})/2$ and $\epsilon_{ij} = \sqrt{\epsilon_{ii}\epsilon_{jj}}$. 

The simple Boltzmann-inversion scheme for the inter-bead potentials that neglects many-body and connectivity effects is approximative. However, the excluded-volume part of the LJ is hardly affected by this treatment, only the attractive (van der Waals) part of the LJ interaction is expected to be affected by the many-body contributions. (Note that most CG simulations in the literature do not include the van der Waals attraction.) Therefore, we tested the influence of varying the bead $\epsilon_{ii}$ on some of the key structural and electrostatic of the dendrimers. The results (presented in the SI) show hardly any influence on the results for dispersion variations in a reasonable window and therefore leave our results quantitatively essentially unchanged. 

\begin{figure}[h]
\includegraphics[scale=0.32]{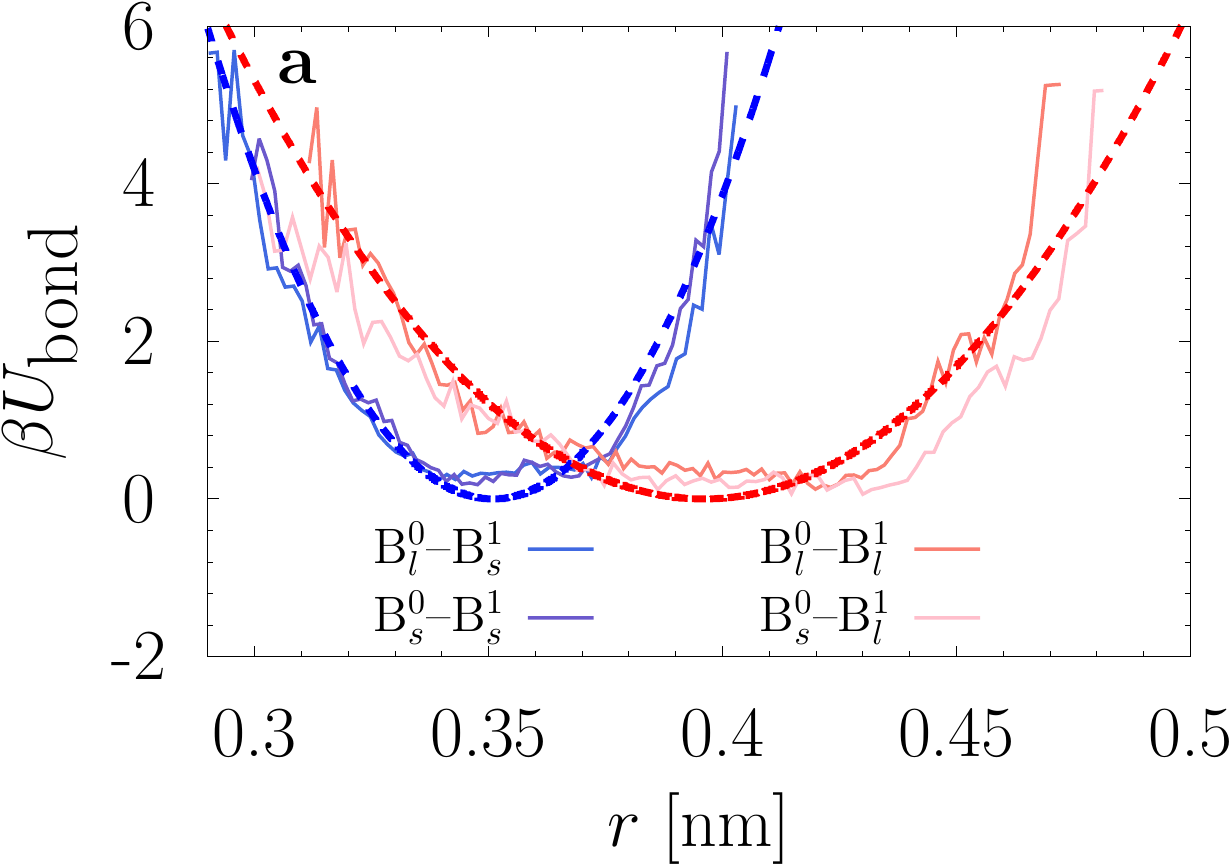}
\includegraphics[scale=0.32]{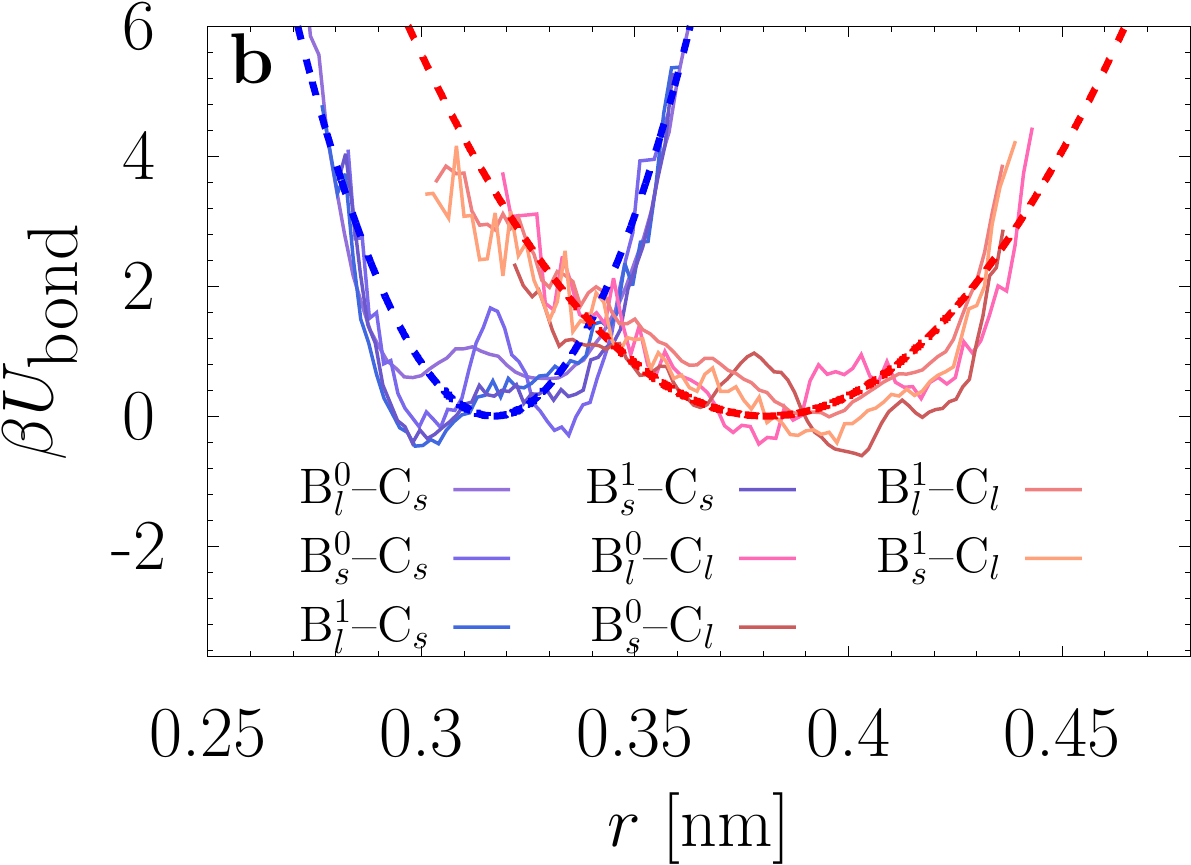}
\includegraphics[scale=0.32]{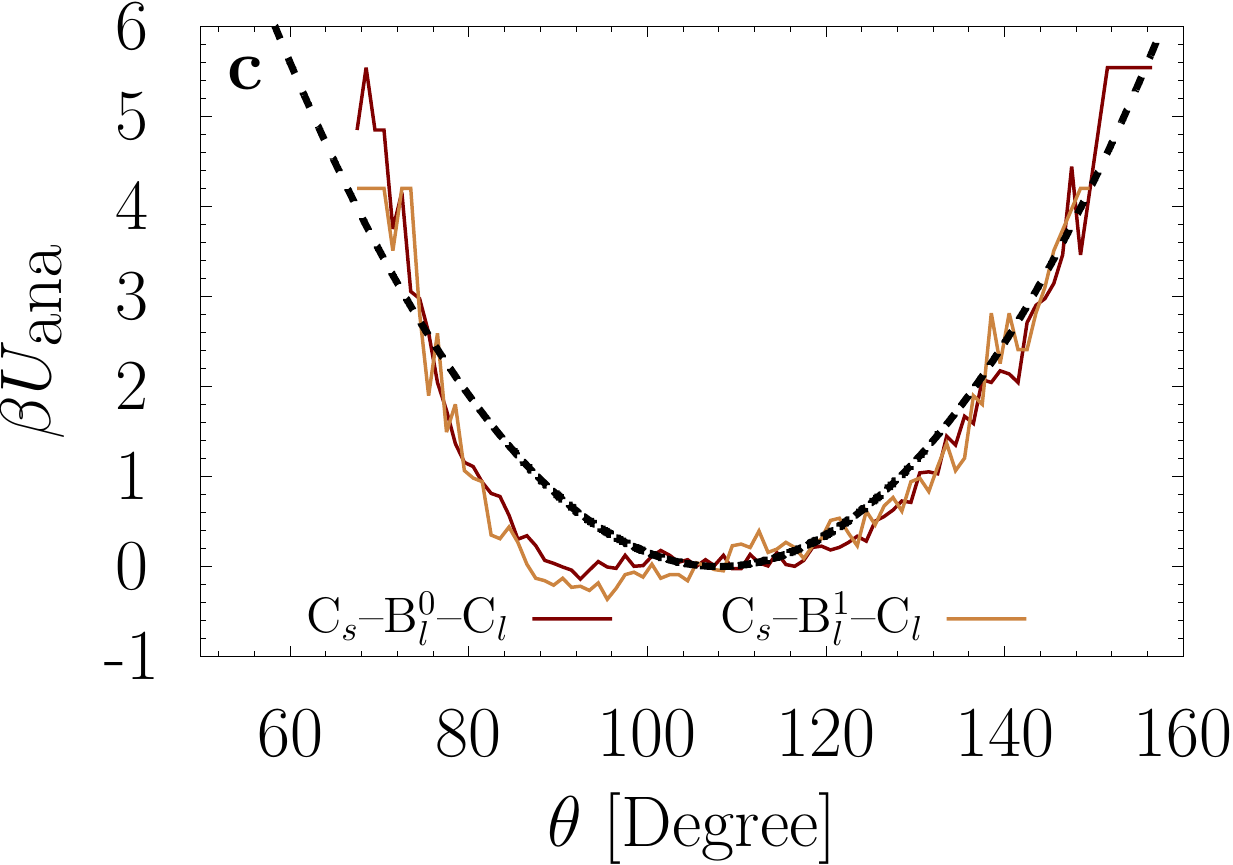}
\includegraphics[scale=0.32]{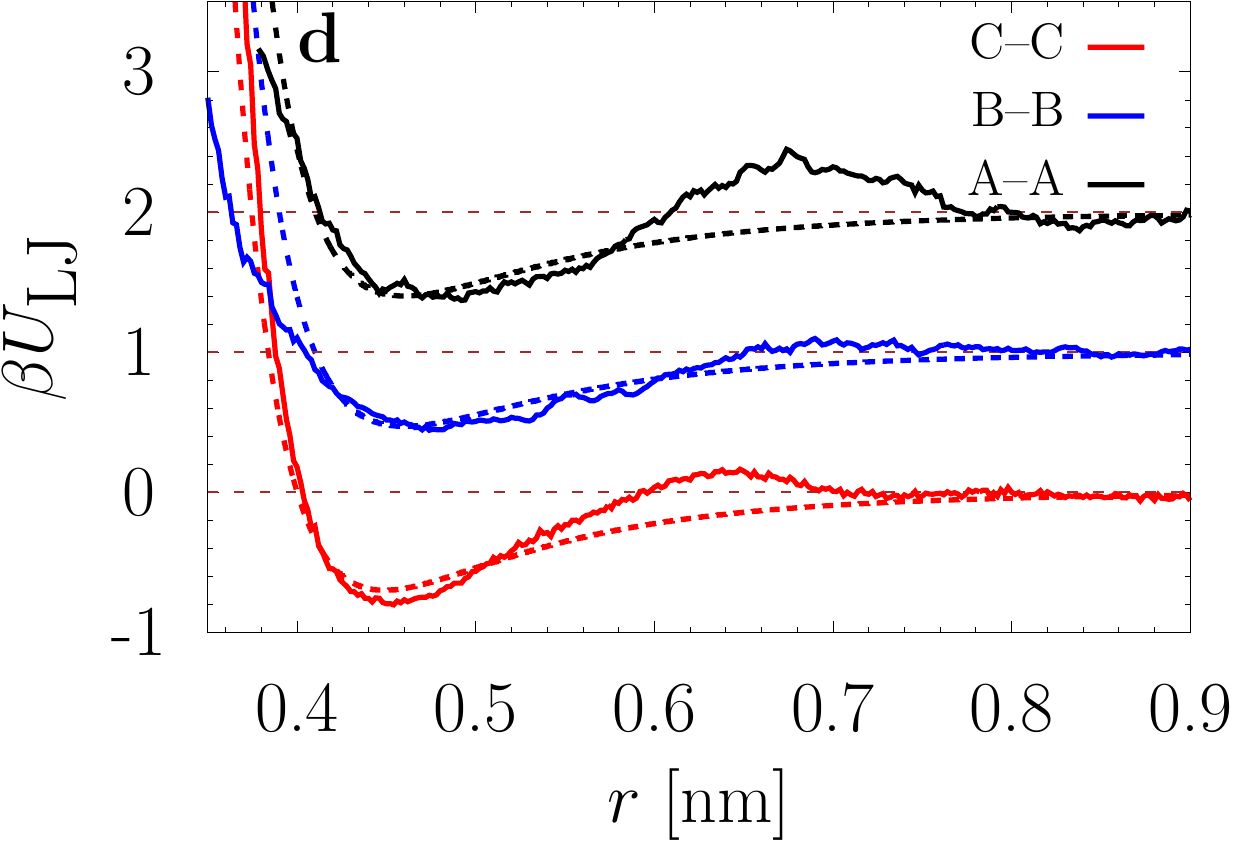}
\includegraphics[scale=0.32]{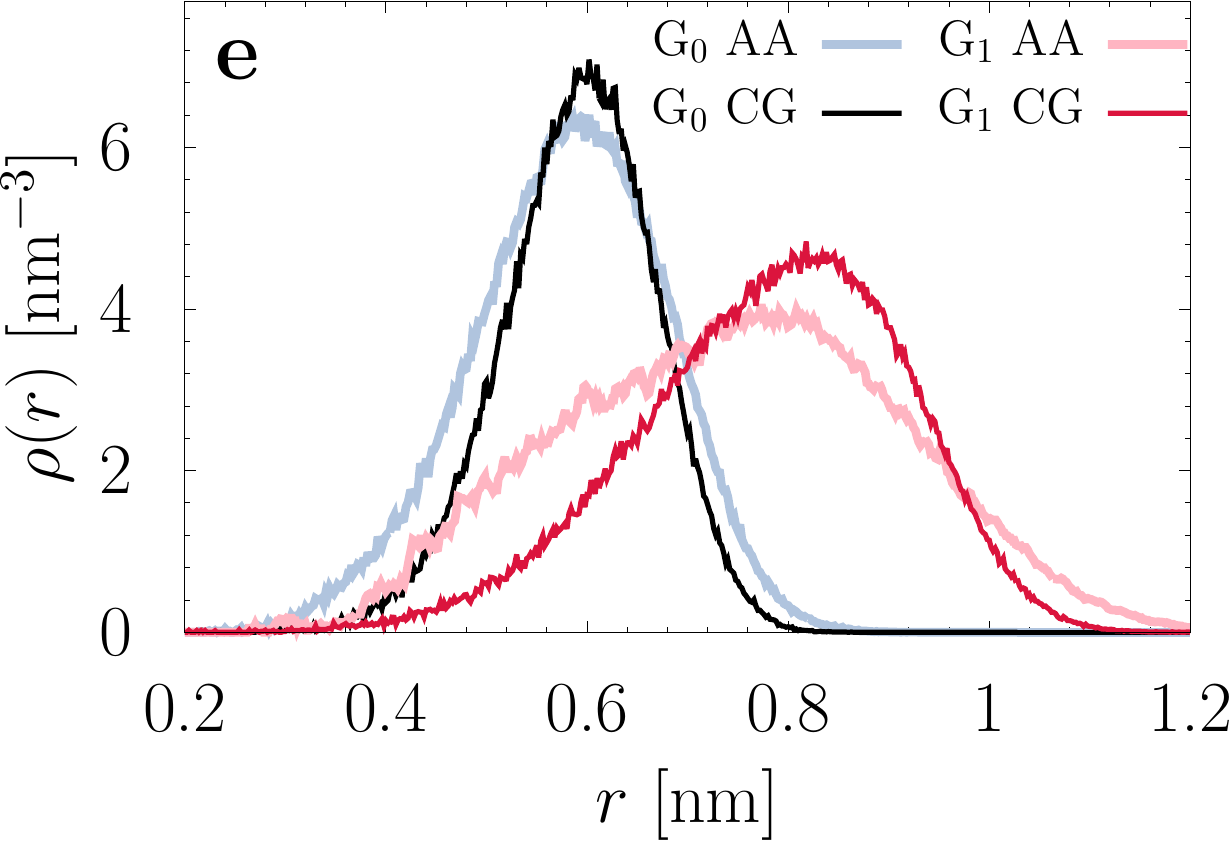}
\includegraphics[scale=0.32]{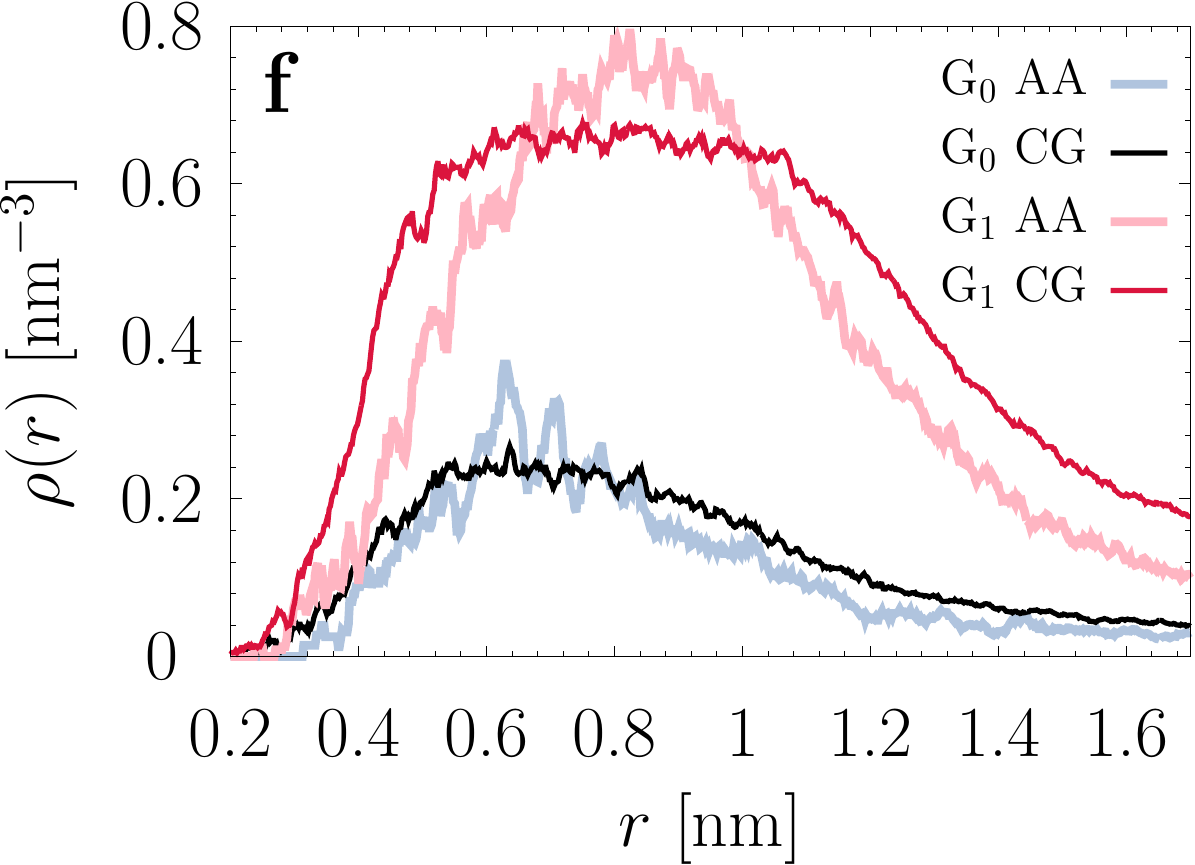}
\caption{The (a) B$_i^{0}$-B$_j^{1}$ bond, (b) B$_i^{m}$-C$_j$ bond, (c) C$_i$-B$_j^{m}$-C$_k$ angular ($m=0$ for G$_0$ and $m=1$ for G$_1$), and (d) A-A/B-B/C-C effective inter-bead potentials extracted from the all-atom MD simulations (AA; solid lines). The dashed lines are the fits according to the coarse-grained  (CG) Hamiltonian Eq.~(\ref{H_intra}).  For the atomistic simulations, the coordinate $r$ is defined as the bead-bead center-of-mass distance. The indices $i,j=s,l$ distinguish between beads having  short or long bond connections in the molecular structure, respectively. 
In the bottom panels the density distributions of the terminal sulfate groups (e) and counterions (f) around the dPGS center-of-mass from atomistic and CG simulations are compared for the G$_0$ and G$_1$ dPGS, respectively.  The simulations were performed at salt concentrations $c = 30$ mM for G$_0$ and $c = 50$ mM for G$_1$.}
\label{E_intra}
\end{figure}

\subsection{Coarse-grained potentials}

Selected results of our mapping procedure are plotted in Fig.~\ref{E_intra}. As we see in panels (a) to (c), the intra-bond potentials can be fitted well by a harmonic function.  We find that the asymmetry in the glycerol repeat unit leads to an equilibrium bond length that differs if `short' or `long' beads are connected at the upper cycle branch $m+1$.  The shift between the bond lengths corresponds to a single covalent bond length on the \AA ngstrom scale.  The structure asymmetry is also reflected in the angular potential $U_{\rm ana}$ (provided in the SI), although in that case the effect is less notable as compared to the bond potential. As shown in Fig.~\ref{E_intra}(c), the angle $\theta$ formed by a triplet of monomers C--B--C has a distribution ranging from $60^\circ$ to $150^\circ$ and is thus relatively broad when compared to typical atomistic potentials. Note that in all of the previous CG dendrimers models angular potentials were typically neglected.
 All bonded bead potentials are summarized in Table~1.

In Fig.~\ref{E_intra}(d), we present the non-bonded LJ potentials between pairwise groups A-A, B-B, and C-C. 
At small separation, $r \lesssim 0.4$ nm, a strong repulsion demonstrates the excluded--volume interaction, while in the intermediate distance $r \simeq 0.6$ to 0.7 nm, a small energy barrier signifies effects of the first hydration shell around the beads.  We neglect these small barriers and fit the curve via the LJ potentials, cf. the dashed lines in Fig.~\ref{E_intra} (d). The final parameters for the CG $\epsilon_{ii}$ and $\sigma_{ii}$ values for all three bead types are very similar. This can be seen as a  justification to set a generic LJ potential for all CG beads as typically utilized in previous CG studies of monomer-resolved dendrimers~\cite{Ballauff2004, Huismann2010, Huismann2010B, Das2014}.
We complement the CG force-field by effective potentials for a symmetric monovalent salt where, for simplicity, we chose the same LJ parameters as for the charged C bead. The corresponding LJ size for the ions is actually  the same as for the effective  CG sodium-chloride cross interaction in water~\cite{Smith1993, Kalcher2010}, so it seems a reasonable choice to model a simple monovalent salt. The relatively small energy parameter $\epsilon_{\rm LJ}$ parameter of 0.1~kJ/mol models hard-sphere like ions without any strong ion pairing behavior.~\cite{Kalcher2010} The parameters for the non-bonded interaction are  summarized in Table~\ref{dPGS_nonbonded}. 

In order to further scrutinize the validity of the CG force field, we compare density profiles of the terminal sulfate groups, Fig.~\ref{E_intra}(e),  and the cations, Fig.~\ref{E_intra}(f), extracted from atomistic and CG simulations (see next section for methods) of generation $n=0$ and 1.  A good agreement is reached between the two approaches in both profiles for both generations, consolidating earlier conclusion on the validity of CG force field for charged dendrimers~\cite{Huismann2012}.  For G$_1$, we find that the width of the distribution of the terminal groups appears to be a bit narrower in the CG force field.  Yet, the global structure represented by the location of the density peaks appears to be similar for both approaches. This similarity applies to the counterion distribution as well, indicated by the density peak at $r \sim 0.82$~nm produced by both levels of modeling. 

\begin{table}[htbp]
\centering
\caption{Parameter for the CG dPGS bonded potentials. The subscript $s$ and $l$ refers to B$_i$ beads with a long or short glycerol arm, respectively. $m=0$ to $n$ denominates the intermediate branching cycle for a given generation $n$.}
\begin{tabular}{p{2.25cm}||P{1.32cm}|P{0.7cm}||P{1.4cm}|P{0.5cm} }
\hline
       & $k_{b}$    & $l_0$     & $k_{a}$  & $\theta$  \\
group &{\small ($\kB T$\,nm$^{-2}$)} &{\small (nm)}     &{\small ($\kB T$\,deg$^{-2}$)}  &{\small (deg)}    \\
\hline
A--{B$_l^0$}       & 1381 & 0.380  & --  & --   \\
A--{B$_s^0$}       & 5100 & 0.337  & --  & --   \\
\hline
B$_{s,l}^m$--B$_l^{m+1}$  & 1052 & 0.395  & --  & --  \\
B$_{s,l}^m$--B$_s^{m+1}$  & 3105 & 0.351  & --  & --   \\
\hline 
{B$_{s,l}^n$}--{C$_l$}    & 1633 & 0.380  & -- & --     \\
{B$_{s,l}^n$}--{C$_s$}    & 6160 & 0.312  & -- & --   \\
\hline
B$_{s,l}^0$--A--\,B$_{s,l}^0$ & --  & -- & 0.003 & 115    \\
B$_{s,l}^1$--B$_{s,l}^0$--A & --        & -- & 0.003 & 115     \\
B$_{s,l}$--B$_{s,l}$--B$_{s,l}$ & -- & --  & 0.003 & 115    \\
C$_{s,l}$--B$_{s,l}^{n}$--B$_{s,l}^{n-1}$ & --   & -- & 0.003 & 121    \\
C$_{s,l}$--B$_{s,l}^0$--A  & --   & -- & 0.003 & 121    \\
C$_{s}$\,--\,\,B$_{s,l}^n$--\,C$_{l}$    & -- & --  & 0.005 & 108  \\
\hline
\end{tabular}
\label{dPGS_bonded}
\end{table}

\begin{table}[htbp]
\centering
\caption{CG dPGS nonbonded potential.}
\begin{tabular}{p{2.3cm}||P{2.3cm}|P{2.3cm}}
\hline
group & $\sigma_{\rm LJ}$ (nm)             & $\epsilon_{\rm LJ}$ ($\kB$T)       \\
\hline
A          & 0.41           & 0.60      \\
B          & 0.41           & 0.53      \\
C          & 0.40           & 0.70     \\
cation     & 0.40           & 0.10     \\
anion      & 0.40           & 0.10     \\
\hline
\end{tabular}
\label{dPGS_nonbonded}
\end{table}

\subsection{Coarse-grained simulations}

Having established a CG force field, all generations $n$ of dPGS can be now readily constructed. Higher generations G$_n$ with index $n > 0$ are created by iteratively bonding two extra glycerol units to the original one (dendritic segment) on the outer shell of generation G$_{n-1}$. This Cayley tree-like~\cite{Voit1995} structure gives an exponentially growing of the polymerization $N_g = 3(2^{n+1}-1) + 1$ with its generation index $n$, which thereby leads to the sulfate group number $N_{\rm ter} = 6(2^{n+1} - 2^{n})$ and the gross number $N_{\rm dP} = 6 \times 2^{n+1} -2$ of the CG segments.
Note the above structure only fits to a perfect dendrimer,  whereas an imperfect dendrimer bears a small number of linear segments that corresponds to most experimental dPGS realizations~\cite{JensDernedde2010}. In our work we investigate seven different generations $n=0$ up to $n=6$, depicted and with some features summarized in Table~\ref{dPGS_para}.

For the CG simulations the dPGS macromolecules are placed in a cubic box with a side length of $L = 35$~nm with periodic boundary conditions in all three directions. We treat the solvent implicitly via a uniform dielectric background, however, ions are treated explicitly to account for ionic correlation effects.  In view of the charge status of dPGS, a number of monovalent counterions, $n_c = 6(2^{n+1} - 2^{n})$, is added to maintain an electroneutral system. Apart from the counterions, in all simulations, dPGS is immersed in a salt solution with $N_i = 257$ to 5140 pairs of monovalent cations and anions, resulting into  bulk salt concentrations from $c_0 = N_i/L^3 \simeq 10$ mM to 200 mM. 
The initial dendrimer configuration is assembled according to the equilibrium bond length $l_0$ and angle $\theta_0$ appeared in Eq.~(\ref{H_intra}). 

We perform Langevin dynamics simulations on CG dPGS of generation number ranging from 0 to 6 also using the GROMACS package. All the implicit water simulations used the second-order stochastic dynamics (SD) integrator in GROMACS with the friction in the Langevin equation set with a time constant of $\tau_t = 1$~ps and integration time of 2~fs. We set all CG beads to have a small mass of $m_i = 0.5$ amu to decrease inertial effects and lower the intrinsic viscosity (i.e., internal relaxation time) of the dPGS. Equilibrium properties, as investigated in this work, are not affected by any reasonable mass choices as long as the simulations are ergodic.
With an increasing number of the terminal beads the electrostatic interaction becomes more profound and the cut-off radius for the PME summation and short-range van der Waals interactions is extended to $r_\textrm{cut}=4$ nm as compared to the above atomistic simulation.  The choice of the cutoff is verified by reference simulations with increased cutoff value $r_\textrm{cut}=6.0$~nm.  Unless specifically stated otherwise, the temperature was set to 310~K as the default. The static dielectric constant of the solvent is  $\epsilon_r = 78.2$ at this temperature. After energy minimization of the initial structure and a 1~ns equilibration period, the production run of a $NVT$ simulation lasts around 60~ns.  
That time has been proven to be sufficient for equilibrium sampling for all generations as in detail verified in the SI by scrutinizing relaxation times and time unit scalings between all-atom and CG simulations.

\subsection{Analysis of the CG simulations} 

\subsubsection{dPGS radius of gyration and asphericity}

The size of a dendrimer can be characterized by the radius of gyration $R_g$ which is defined as the trace of the gyration tensor (${\bold R}^2$)~\cite{Carbone2010}
\begin{equation}
{\bold R_{\alpha,\beta}}^2 = \frac{1}{N_{\rm dP}}\sum^{N_{\rm dP}}_{i=1} ({\bold r}^i_{\alpha} - {\bold r}^M_{\alpha}) ({\bold r}^i_{\beta} - {\bold r}^M_{\beta}), \,\,\, \alpha, \beta = x,y,z,
\end{equation}
where ${\bold r}^i_{\alpha}$ is the coordinate of the $i$th segment and ${\bold r}^M_{\alpha}$ is the dPGS COM position along the $\alpha$ direction. The square of the radius of gyration is then 
\begin{equation}
{R_g}^2 = \langle tr({\bold R}^2)\rangle = \langle \sum^3_{i=1} {\lambda_i}^2\rangle,  
\end{equation}
where ${\lambda_i}^2$ is the $i$th eigenvalue of the gyration tensor, representing the characteristic length of the equivalent ellipsoid which mimics the dendrimer. $\langle \cdots \rangle$ stands for the ensemble average.
The degree of asphericity $A$ of dPGS is defined in terms of the eigenvalue ${\lambda_i}^2$, which can be written as
\begin{equation}
A = \frac{\langle (T_r^2 - 3M)\rangle}{\langle T_r^2 \rangle},
\label{define_A}
\end{equation}
with $T_r = {\lambda_1}^2 + {\lambda_2}^2 + {\lambda_3}^2$ and $M = {\lambda_1}^2 {\lambda_2}^2 + {\lambda_2}^2 {\lambda_3}^2 + {\lambda_1}^2 {\lambda_3}^2$. Fot a perfect sphere $A$ equals 0, whereas $A = 1$ corresponds to the extreme of an infinitely thin rod.

\subsubsection{The radial electrostatic potential}
The electrostatic potential $\phi$ is available through the framework of Poisson's equation 
\begin{eqnarray}
\nabla ^2 \phi = - {\sum_{i=+,-,C} Z_i \rho_i(r)}/{\epsilon_0 \epsilon_r},
\end{eqnarray}
where in our CG system $\rho_+(r)$, $\rho_-(r)$ and $\rho_{\rm C}(r)$ are the distance--resolved radial density profiles for all charged species, namely cations, anions and sulfate (C) beads, respectively.
We integrate Poisson's equation numerically feeding in the $\rho_i(r)$ generated from simulation to obtain the local electrostatic potential $\phi(r)$.  In addition to the potential profile, we calculate the running coordination number of charged beads of type $i=\pm,C$:
\begin{equation}
N_{i}(r) = \int^r_0 \rho_{i}(s) 4\pi s^2 ds .
\end{equation} 
Since dPGS is negatively charged,
$N_{+}(r)$ explicitly represents for the number of counterions located in a distance $r$ away from the dPGS COM.
It follows that the total accumulated charge is 
\begin{eqnarray}
Z_{\rm acc}(r) = N_{+}(r) - N_{-}(r) - N_{C}(r),
\label{acc}
\end{eqnarray}
which gives the accumulated dPGS charge deduced from the structural one by adding the ionic shell.

For the definition of the effective charge of a dendrimer (and thus the effective surface potential) we take the basic Debye-H\"uckel (DH) theory for the radial electrostatic potential distribution around a charged sphere with radius $r_{\rm eff}$ and valency $Z_{\rm eff}$ as reference~\cite{Belloni1998}, 
\begin{equation}
e \beta \phi_{\rm DH}(r) = {Z_{\rm eff}} l_B \frac{e^{\kappa r_{\rm eff}}}{1 + \kappa r_{\rm eff}} \frac{e^{-\kappa r}}{r},
\label{DH_solution}
\end{equation}
where $e$ is the elementary charge and $\kappa = \sqrt{8\pi \lB c_0}$ is the inverse Debye length for a symmetric, monovalent salt.
This solution is derived with the Dirichlet boundary condition, \ie, fixing the surface potential $\phi(r_{\rm eff})$ and the one far away $\phi(\infty) = 0$.
The DH potential usually works well in the region far from the colloid, 
where nonlinear effects, such as ion--ion correlations and condensation, become irrelevant.   For the ``correlated Stern layer'' at the interface of the charged sphere, the electrostatic potential $\phi$ is expected to  deviate strongly from the DH potential $\phi_{\rm DH}$ and all nonlinear effects are adsorbed into the effective charge $Z_{\rm eff}$ (as, e.g., based on solutions of the full non-linear Poisson-Boltzmann theory~\cite{Ohshima1982, Alexander1983, Ramanath1988, Belloni1998, Bocquet2002}). 
By taking the logarithm of eq.~(\ref{DH_solution}) and mapping directly on the far-field behavior of the electrostatic decay calculated in the simulation, the double-layer behavior can be quantified with high accuracy.~\cite{Kalcher2009} This provides also the basis to define the position of the Stern layer, or better expressed, the exact location $r_{\rm eff}$ of the interface between the diffusive double layer in the DH sense and the correlated condensed ion layers. The effective surface potential is then simply $\phi_0 = \phi_{\rm DH}{(r_{\rm eff})}$. 

\subsection{Experimental Materials and Methods}
\begin{table}[htbp]
\centering
\caption{Properties of dPGS of generation $G_n$ in the experiments.
The dPGS weight $M_{\rm n, dPGS}$ is deduced from the respective core weight $M_{\rm n, dPG}$ and sulfate group number $N_{\rm ter}$.
DS is the degree of sulfation, and PDI is the polydispersity index. $\eta$ is the $\zeta$-potential attained from the electrophoretic experiment.}
\begin{tabular}{p{2.0cm}||P{1.1cm}|P{1.1cm}|P{1.1cm}|P{1.1cm}}
\hline
Label          & G$_2$          & G$_4$    & G$_{4.5}$   & G$_{5.5}$   \\
\hline
$M_{\rm n, dPG}$ [kD]  & 2              & 7        & 10          & 20     \\
\hline
PDI            & 1.7            & 1.7      & 1.5         & 1.2    \\
\hline
DS [\%]        & 100            & 100      & 99          & 98     \\
\hline
$N_{\rm ter}$  & 28             & 102      & 135         & 266    \\
\hline
$M_{\rm n, dPGS}$ [kD]   & 5              & 18       & 24          & 47     \\
\hline
$\eta$ [mV] & -47.71  & -58.46   &  -58.73   & -70.9  \\
\hline
\end{tabular}
\label{dpgs_data}
\end{table}

Dendritic polyglycerol (dPG) was synthesized by anionic ring opening polymerization of glycidol~\cite{Sunder1999}. Gel permeation chromatography (GPC) is empolyed to measure the number averaged molecular weight of the core $M_{\rm n,dPG}$ and Polydispersity index (PDI). Afterwards dPGS was prepared by the sulfation of dPG with SO$_3$-pyridine complex in dimethylformamide (DMF) according to reported procedure~\cite{Turk2004}. The degree of sulfation (DS) was determined by elemental analysis. Properties of dPGS in different generations are summarized in Tab.~\ref{dpgs_data}.

Size and $\zeta$-potential (electrophoretic mobility) measurements were performed with a Zetasizer Nano ZS instrument (ZEN 3500, Malvern Instruments, Herrenberg, Germany) equipped with a 18~mW He-Ne laser ($\lambda$=633~nm). The molecule size $r_{\rm hd}$ was measured by dynamic light scattering (DLS) in UV-transparent disposable cuvettes (VWR, Germany) at a back scattering angle of 173$^{\circ}$. The compounds were dissolved in 10~mM MOPS buffer (adding NaCl to adjust ionic strength to 10~mM) pH 7.4 at concentration of 1~mg/ml and were filtered through 0.8~$\mu$m polyethersulfone syringe filter (PALL, USA). Prior to measurement, each sample was equilibrated for 2~min at 37$^{\circ}$ and measured with 10 scans each lasting for 10~s. The stated values for the hydrodynamic diameter are the mean of three independent measurements. 
The electrophoretic mobility was measured at 5~mg/ml in the same buffer as above also in three independent measurements (with all values reported in the SI). The solutions were filtered through 0.2~$\mu$m polyethersulfone syringe filter and equilibrated for 10~min at 37$^{\circ}$ in folded DTS 1060 capillary cells (Malvern, UK). The shown data for the $\zeta$-potential in the resulting figure are based on the Henry function with the Ohshima approximation 
\begin{equation}
f(\kappa r_{\rm hd}) = 1 + \frac{1}{2[1+\delta/(\kappa r_{\rm hd})]^3},
\label{Oappro}
\end{equation}
with $\delta = (5/2)[1+2\exp(-\kappa r_{\rm hd})]^{-1}$ for the conversion of mobilities to potentals~\cite{Ohshima1994}. 
The following reported $\zeta$-potential are the mean of the three independent measurements. 




\section{Results and Discussion}

\begin{figure}[h]
\includegraphics[scale=0.62]{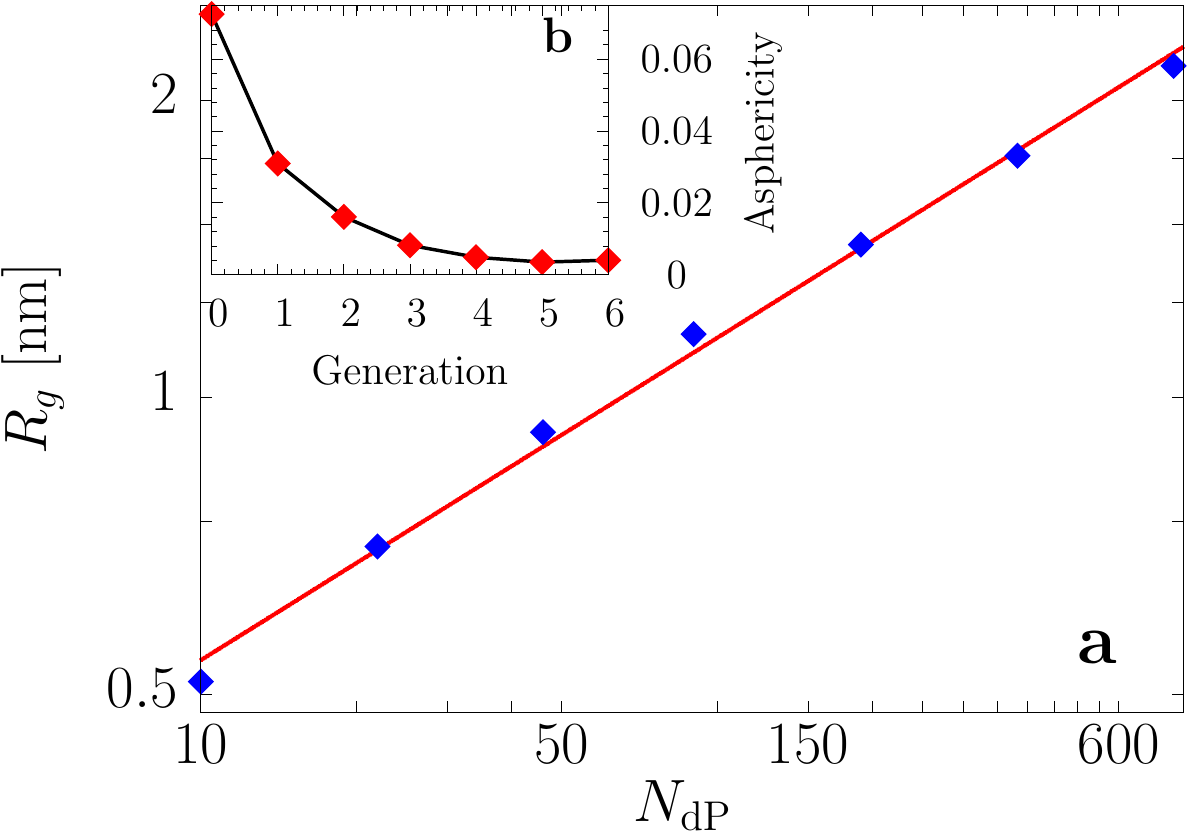}
\caption{ (a) Log--log plot of the radius of gyration $R_g$ versus the total beads number $N_{\rm dP}$ for generations G$_0$ to G$_6$.  The red line is the fitted power law function $R_g \sim N^{0.33}_{\rm dP}$ for all generations, while the dotted cyan line is a fitted power law function $R_g \sim N^{0.30}_{\rm dP}$ for generations $G_2$ to $G_6$ only.
Inset (b): the asphericity parameter $A$ (bottom panel) versus generation of the CG dPGS molecules.}
\label{asph}
\end{figure}

\subsection{Size, sphericity and molecular density distributions}
\begin{figure}[h]
\includegraphics[scale=0.6]{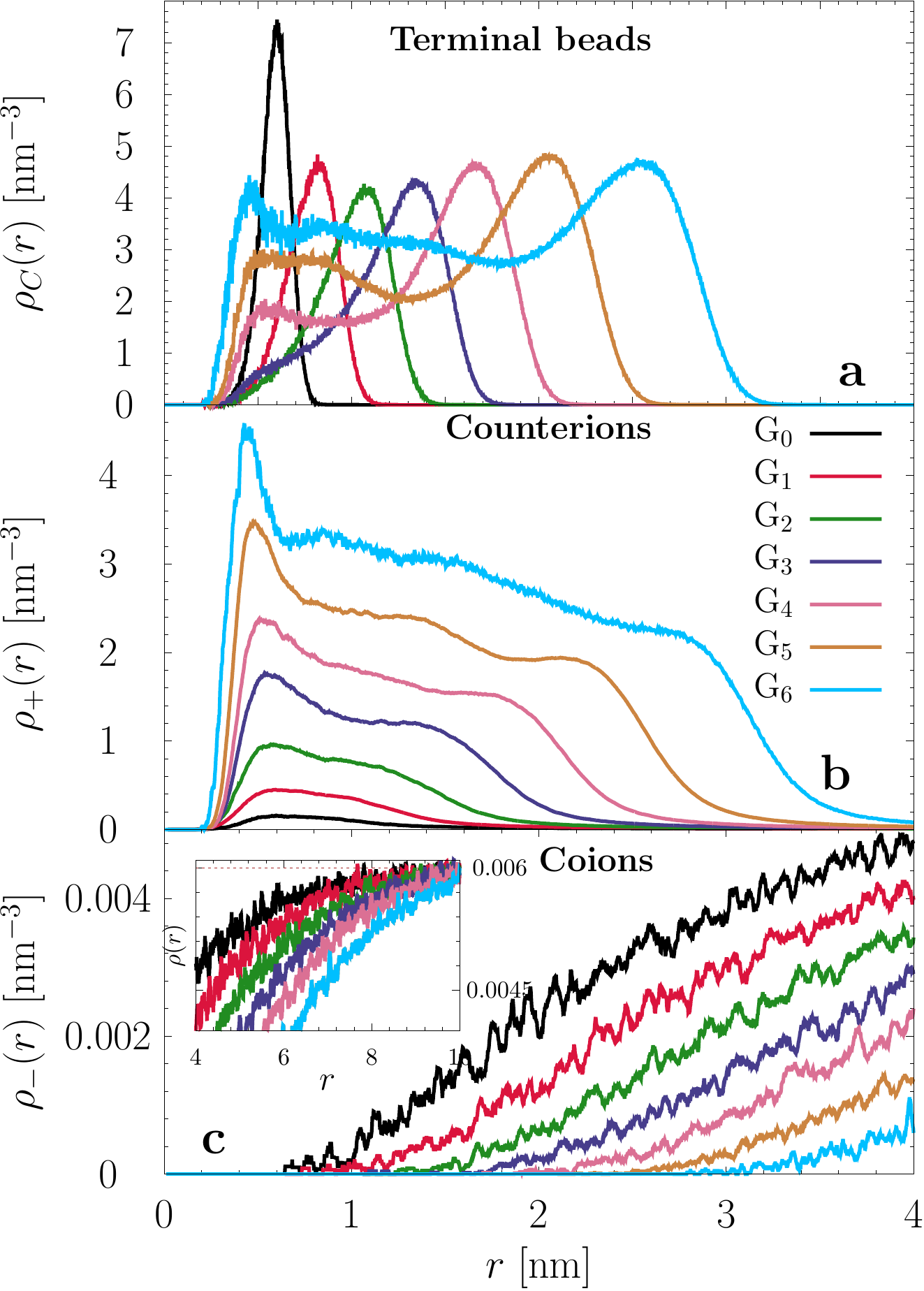}
\caption{
Radial density distribution $\rho_i(r)$ with respect to the dPGS core of (a) the terminal sulfate beads, $i=$C, (b) the cations (counterions; $i=+$),  and (c) the anions (coions; $i=-$) for generations G$_0$ to G$_6$. 
In the inset of panel (c) we show the anion density profile at longer distance up to $r = 10$~nm, 
with the limiting bulk anion density $c_0$ marked by a dashed horizontal line.
}
\label{rdf_dpgs}
\end{figure}

\begin{table*}[htbp]
\centering
\caption{Snapshots and Parameters for the CG dPGS model.
In the dPGS snapshots on the top, red and orange beads depict the terminal charged and inner neutral beads, respectively.
$N_{\rm dP}$ and $N_{\rm ter}$ stand for the total number of CG segments and terminal sulfate (C) beads for dPGS, respectively.
We assign $Z_{\rm bar} = -N_{\rm ter}$, $r_d$ and $R_g$ as dPGS bare charge, radius, and radius of gyration, respectively.
$Z_{\rm eff}$ and $r_{\rm eff}$ define the dPGS effective charge and corresponding radius. Via the inflection point criterion,~\cite{Belloni1998, DavidA.J.Gillespie2014} we can calculate the inflection dPGS radius $r_{\rm inf}$ and accordingly the inflection dPGS charge $Z_{\rm inf}$.
Finally, $\sigma_d = Z_{\rm bar}/(4\pi r^2_d)$, $\sigma_{\rm eff} = Z_{\rm eff}/(4\pi r^2_{\rm eff})$ and $\sigma_{\rm inf} = Z_{\rm inf}/(4\pi r^2_{\rm inf})$ denote the bare, effective surface charge density, and inflection surface charge density, respectively.
At the dPGS surface, we assign $\phi(r_{\rm eff})$ as the surface potential.
All the estimates are made given a salt concentration of $c_0 = 10$~mM.
 }
\begin{tabular}{P{2.3cm}|P{1.7cm}|P{1.7cm}|P{1.7cm}|P{1.7cm}|P{1.7cm}|P{1.7cm}|P{1.7cm}|}
 & \includegraphics[scale=0.05, bb=0 -160 355 324]{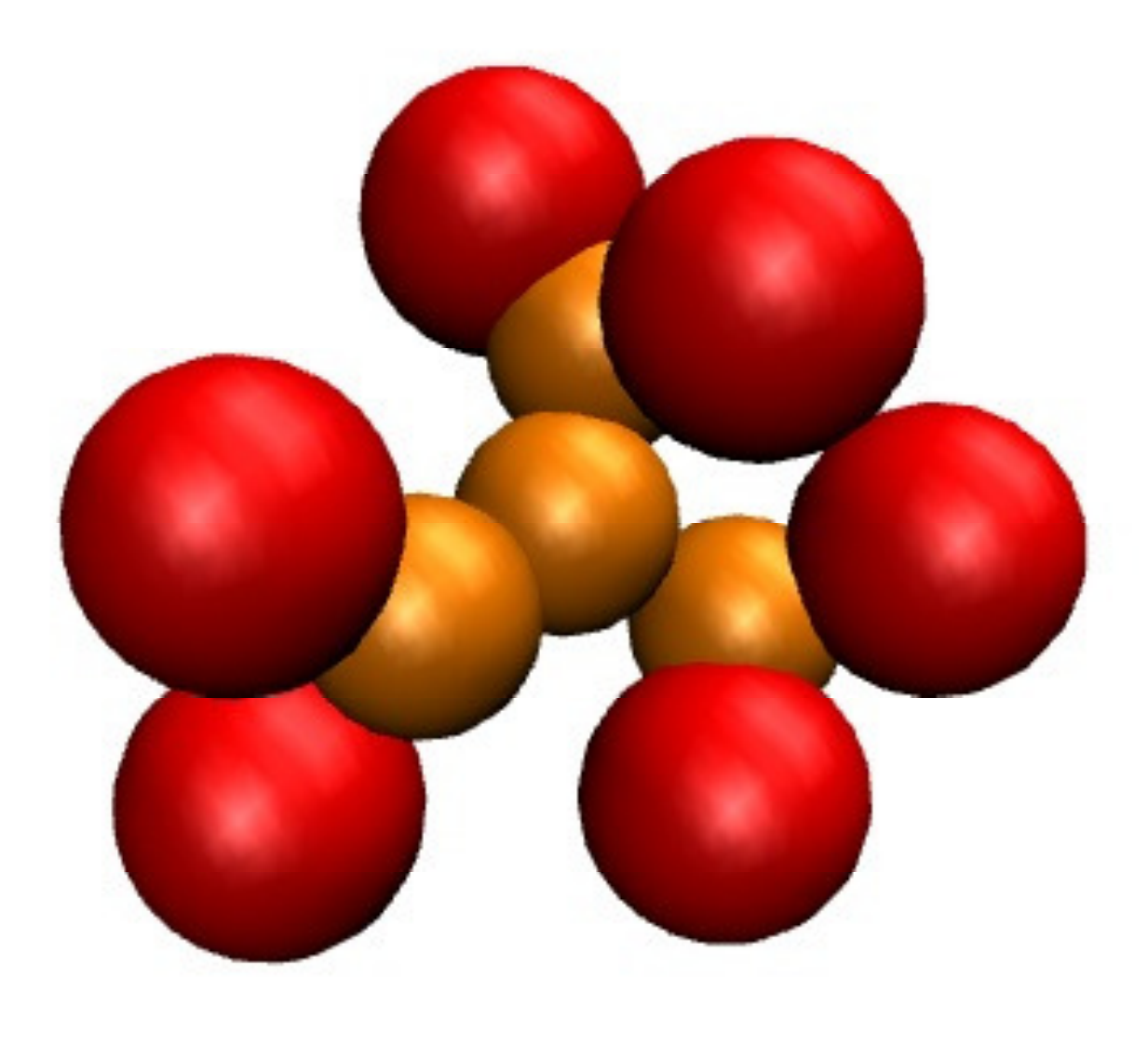} & \includegraphics[scale=0.04, bb= 0 -170 472 498]{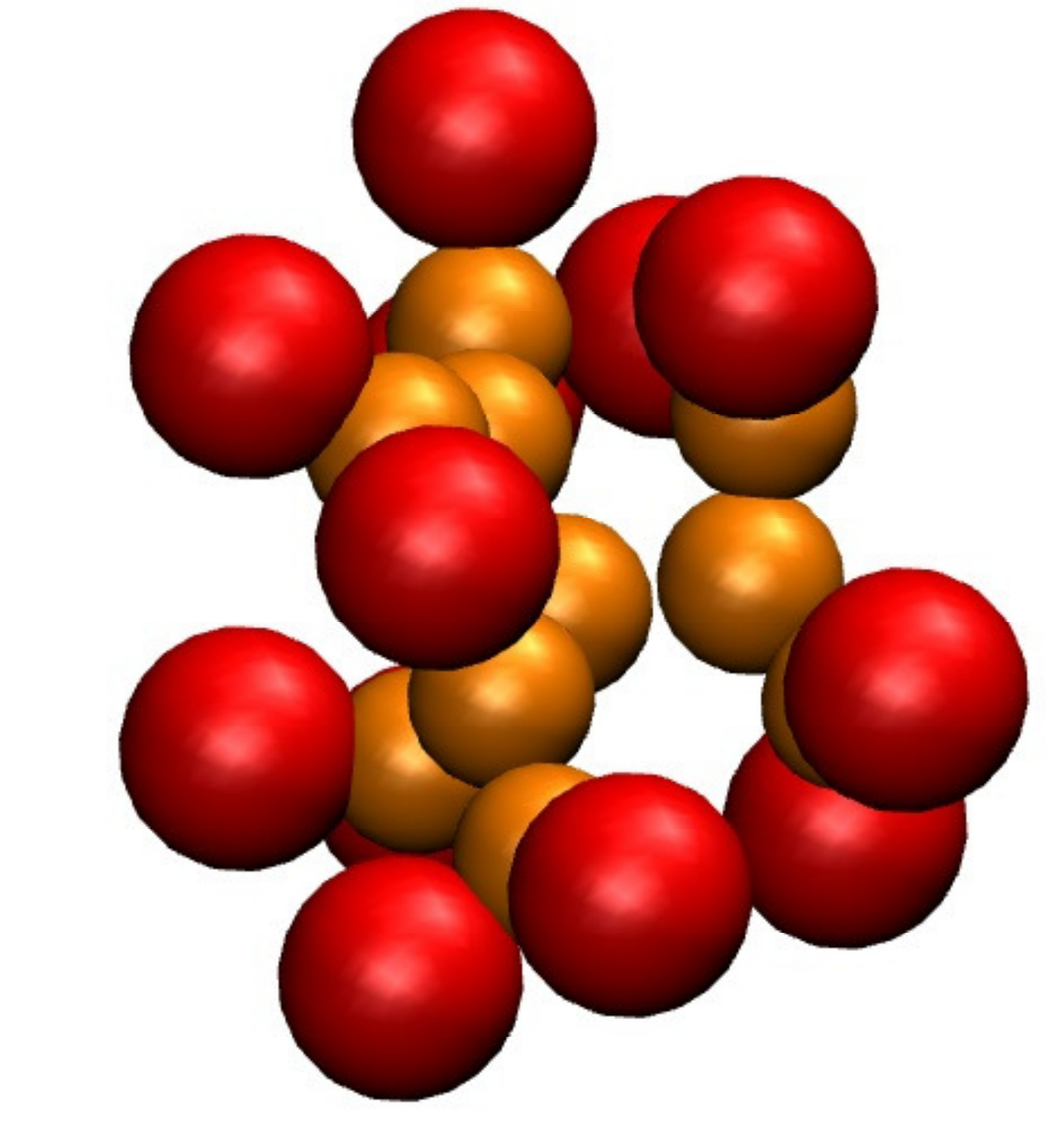} & \includegraphics[scale=0.1]{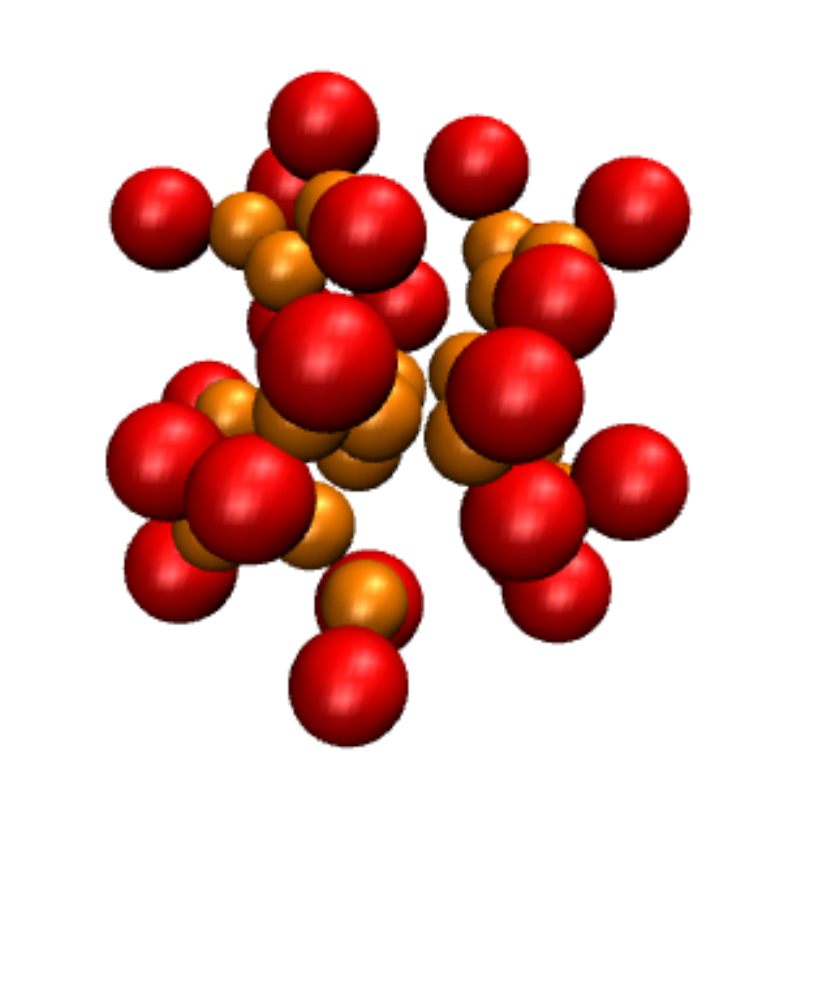} & \includegraphics[scale=0.15]{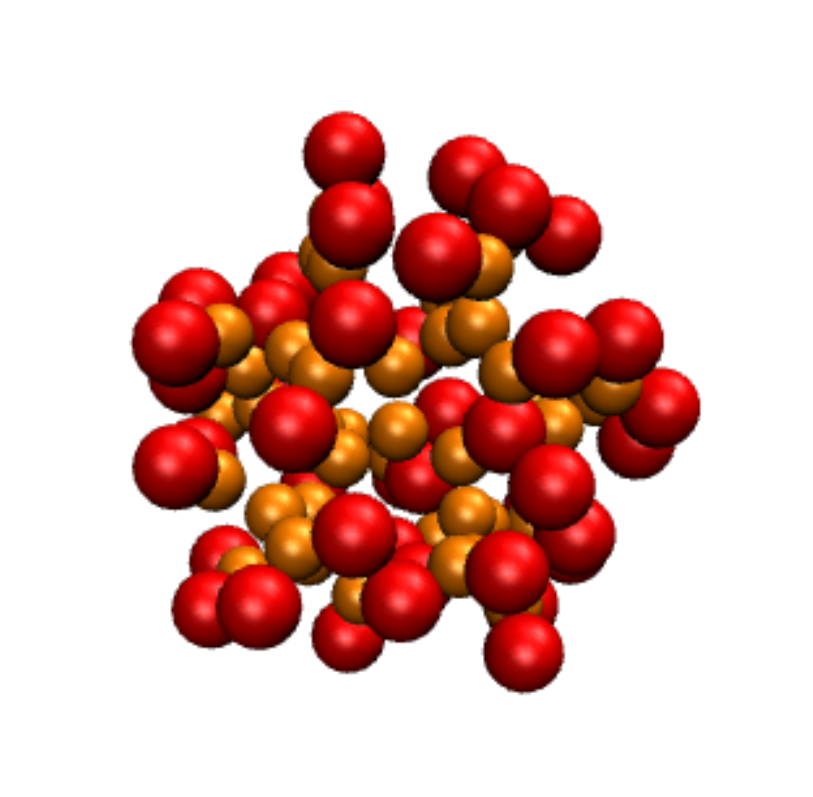} & \includegraphics[scale=0.15]{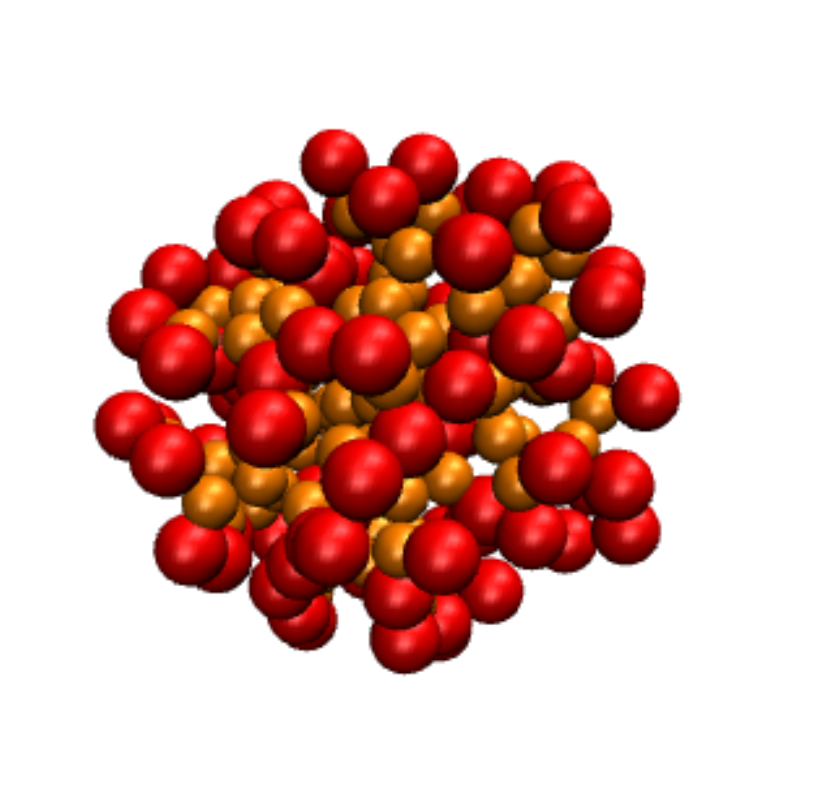} & \includegraphics[scale=0.15]{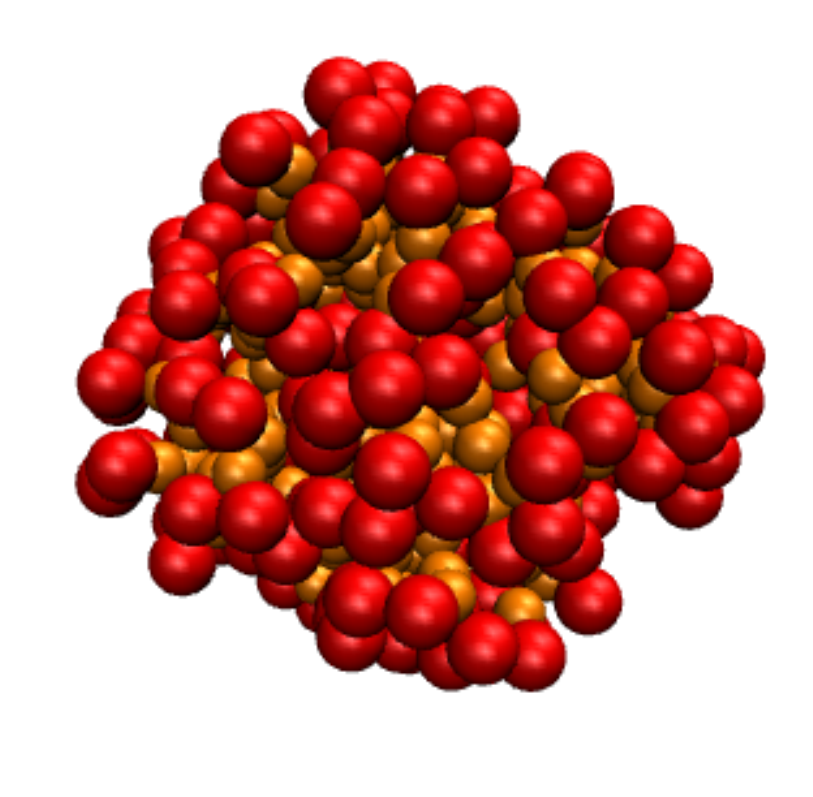}  & \includegraphics[scale=0.15]{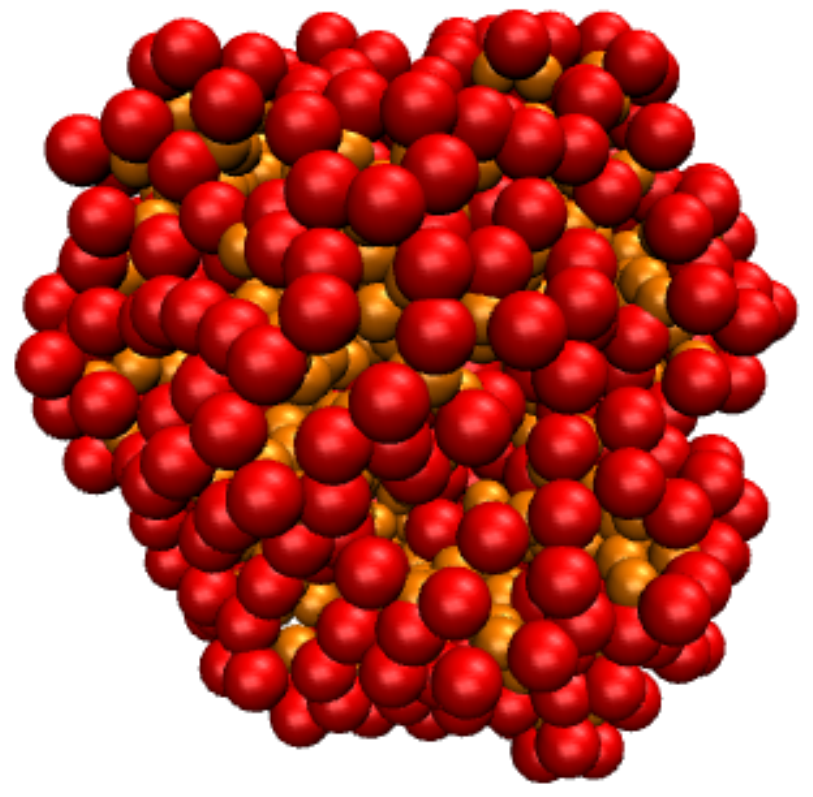}   \\
\hline
\hline
Label & G$_0$ & G$_1$ & G$_2$   & G$_3$  & G$_4$  & G$_5$  &  G$_6$    \\
\hline
MW [KDa] &0.79 & 1.72   & 4.10    &  8.32   & 16.77   & 33.67  &68.00   \\
\hline
$N_{\rm dP}$ & 10 & 22  & 46  & 94   &  190   &  382   &  766  \\
\hline
$N_{\rm ter}$ & 6  & 12 & 24    &  48   &   96    &  192    & 384    \\
\hline
$Z_{\rm bar}$~[$e$] & -6 & -12 & -24  & -48   &  -96  & -192   & -384  \\
\hline
$R_g$~[nm] & 0.52 & 0.71 & 0.92  &  1.16   & 1.43   & 1.76  &2.17   \\
\hline
$r_d$~[nm] & 0.65 & 0.83 & 1.10 &  1.35   & 1.65   & 2.05  &2.55   \\
\hline
$\sigma_d$~[$e$/nm$^{-2}$] & -1.28 & -1.34 & -1.58    & -2.10    & -2.81   & -3.64  & -4.70   \\
\hline
$r_{\rm eff}$~[nm] & 0.7 & 1.6 & 1.9  & 2.4  & 2.8  & 3.3  & 3.8    \\
\hline
$Z_{\rm eff}$~[$e$] & -6.0 & -7.3 & -10.6 & -14.3 & -18.7  & -24.5 & -32.9  \\
\hline
$\sigma_{\rm eff}$~[$e$/nm$^{-2}$] & -0.97 & -0.23 & -0.23 & -0.20 & -0.19  & -0.18 & -0.18  \\
\hline
$\phi(r_{\rm eff})$~[$\kB T$] & -4.20 &-2.12 & -2.37 & -2.22 & -2.28  & -2.25 & -2.40  \\
\hline
$r_{\rm inf}$~[nm] & -- & 1.1 & 1.5   &  1.8   & 2.1   & 2.6  & 3.1   \\
\hline
$Z_{\rm inf}$~[$e$] & -- & -9.9  & -14.9   &  -22.9   & -37.5   & -52.2  & -85.8   \\
\hline
$\sigma_{\rm inf}$~[$e$/nm$^{-2}$] & -- & -0.65 & -0.53   &  -0.56   & -0.68   & -0.62  & -0.71   \\
\hline
\end{tabular}
\label{dPGS_para}
\end{table*}

\subsubsection{Radius of gyration and asphericity}

The radius of gyration $R_g$ of the dPGS macromolecules as a function of generation $n$ is summarized in Table~\ref{dPGS_para}.  We find that $R_g$ increases from $0.52$ nm to $2.17$ nm from G$_0$ to G$_6$. A linear behavior is revealed in a log--log plot of $R_g$ in terms of the total number of the CG segments,  $N_{\rm dP}$, in Fig~\ref{asph}(a). 
Hence, the dPGS size follows the scaling law $R_g \sim N^{\nu}_{\rm dP}$, where we find the scaling exponent $\nu = 0.33$ if we fit all generations, while it decreases slightly to $\nu = 0.30$ if we only fit through the larger generations $G_2$ to $G_6$, cf. Fig~\ref{asph}(a). 
Such a scaling close to 1/3 is fully consistent with the now well established 'dense-core' picture of dendrimers, where details, however, can depend on the particular dendrimer architecture, see the deeper discussions in exemplary previous work.~\cite{Ballauff2004,Tian2013,Maiti2009,Klos2013}.  

In Fig~\ref{asph}(b), we plot the asphericity versus the generation number. For all inspected generations we find values lower than ${\rm A}\sim0.07$, which suggests an almost perfect spherical shape for dPGS molecules. Larger generations show higher sphericity, very likely due to a more homogeneous distribution of the larger number of closer packed beads and a thus higher compactness.   We show snapshots of the CG dendrimers for all investigated generations G$_0$ to G$_6$ in Table~\ref{dPGS_para}.

\subsubsection{Density distributions and `intrinsic' radius}

Figure~\ref{rdf_dpgs} shows the radial density distributions $\rho_i(r)$ of selected components with respect to the distance $r$ to the dendrimer core bead. Fig~\ref{rdf_dpgs}(a) shows the distribution of the terminal sulfate beads. 
For the smaller generations $G<4$, we find a single-peaked distribution, corresponding to the picture that most of the charged terminal beads stay on the molecular surface~\cite{Huismann2010,Huismann2010B, Klos2010, Klos2013}. For the larger generations, however, a bimodal distribution signified by a small peak at $r\simeq 0.6$~nm appears, indicating a small number of dendrons backfolding toward the dendrimer core. (Better visible in density plots re-scaled to refer to the position of the sulfate peak, see the SI.)  The backfolding effect was already detected for other terminally-charged CG dendrimer models~\cite{Huismann2010,Huismann2010B,Klos2010, Klos2013} contributing to a dense-core in contrast to a dense-shell arrangement.~\cite{Ballauff2004}  An `intrinsic' dPGS radius $r_d$ can be roughly deduced from the location of maximum density of the terminal groups. As implied in Fig.~\ref{rdf_dpgs}(a), we find a dPGS radius $r^{G_0}_{d} \simeq~0.6$ nm for generation G$_0$ increasing to $r^{G_6}_{d} \simeq 2.6$~nm for generation G$_6$. All values are summarized in Table~\ref{dPGS_para}.

The radial density $\rho_+(r)$ of the cations, shown in  Fig.~\ref{rdf_dpgs}(b), is found to have qualitatively the expected response to the distribution of terminal beads and follows roughly the sulfate distribution.  In particular, for all generations $\rho_+(r)$ decreases as expected in an exponential (Yukawa or DH-like) fashion to the bulk concentration for large distances. For closer distances, $r\simeq r_d$ highly nonlinear effects are visible, e.g., in the response to the backfolding of the terminal beads we observe the enrichment of cations at $r \simeq 0.5$ nm inside the dPGS, corresponding to the lower sulfate peaks close to the core in Fig.~\ref{rdf_dpgs}(a). On the contrary and as expected,  coions are repelled by the dPGS due to the electrostatic repulsion. 

\begin{figure}[h]
\includegraphics[scale=0.65]{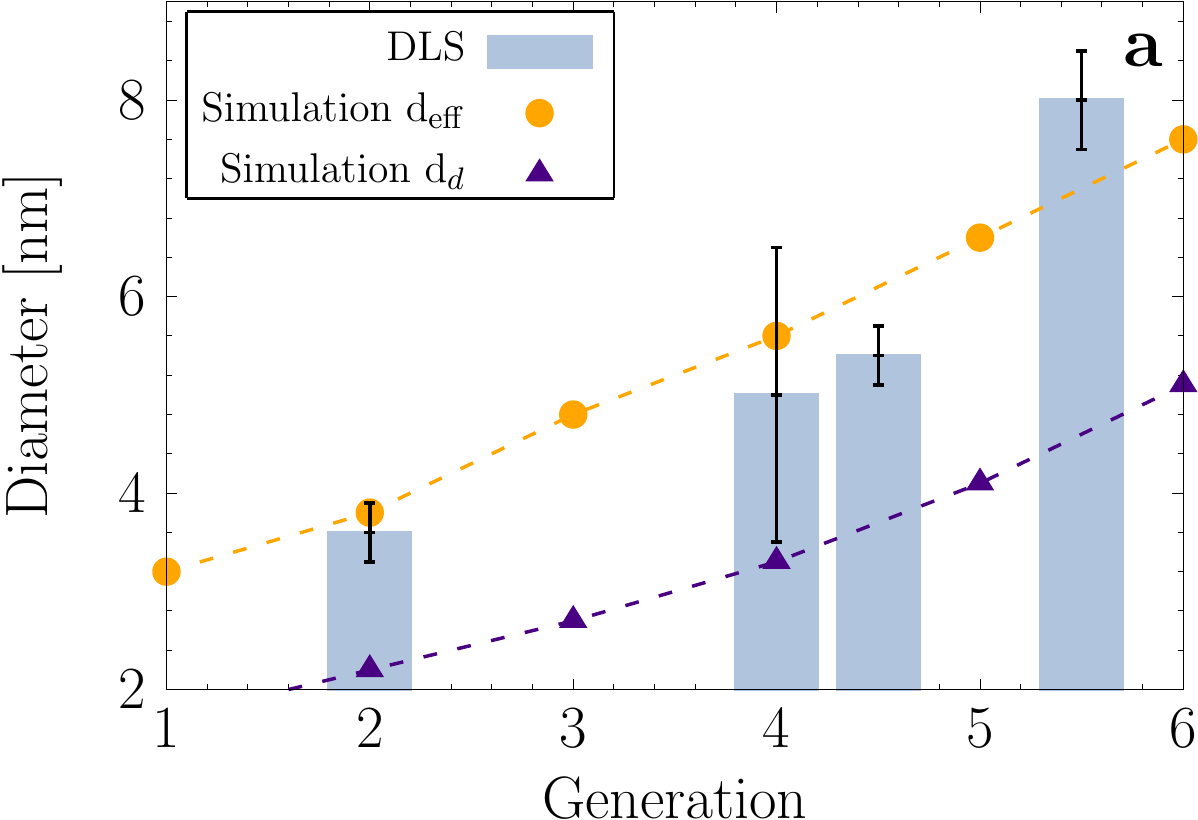}
\includegraphics[scale=0.65]{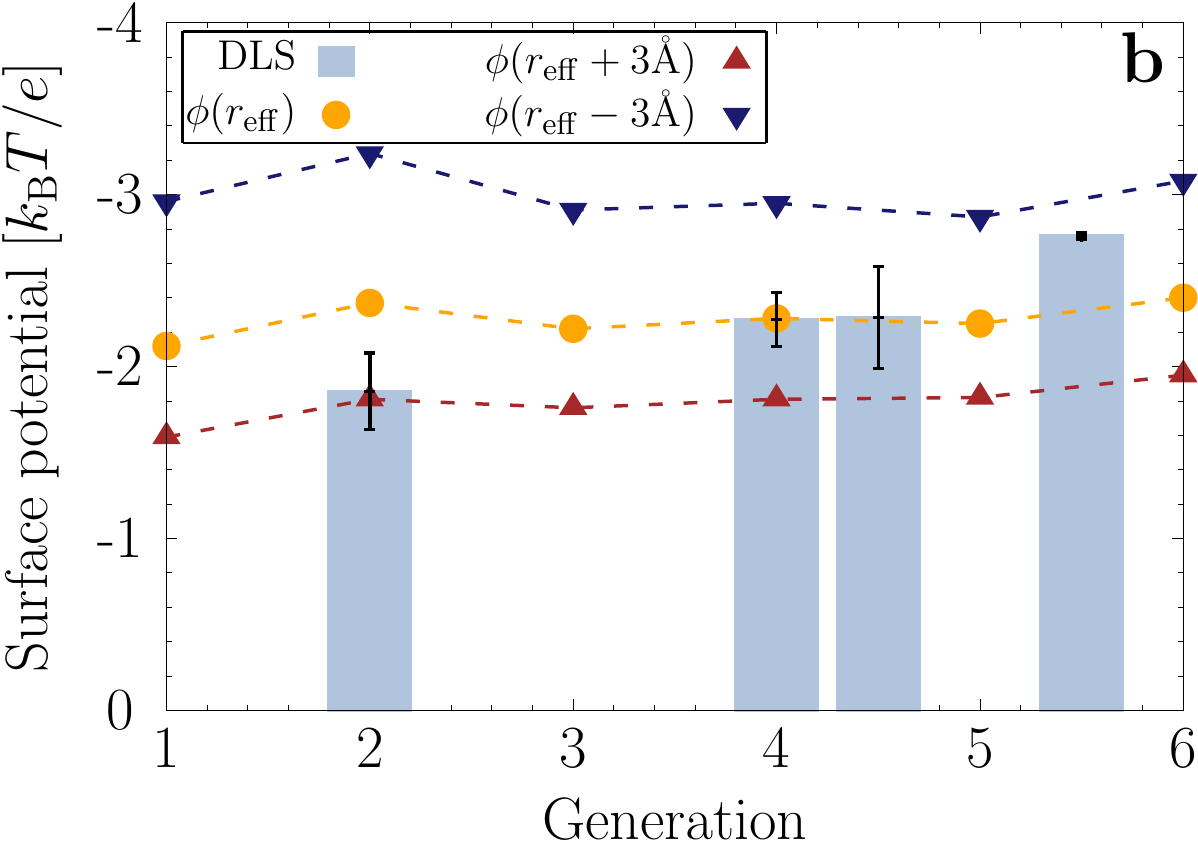}
\caption{
(a) Hydrodynamic diameter and (b) surface potential as determined via Zetasizer measurements (blue bars) and simulations (symbols; dashed lines are guided to the eye). In a) the simulation results are plotted for the intrinsic dPGS diameter $d_d = 2r_d$ and the effective diameter $d_{\rm eff}=2 r_{\rm eff}$. In b) simulation results are plotted for the effective surface potential $\phi_{\rm eff}=\phi(r_{\rm eff})$ at the location $r_{\rm eff}$ and that at one solvation layer shifted, i.e., $r_{\rm eff}+0.3$~nm~and $r_{\rm eff}-0.3$~nm, respectively. }
\label{DLS}
\end{figure}

In Fig~\ref{DLS}(a), we plot the hydrodynamic diameter measured via Zetasizer experiments of generation 2, 4, 4.5 and 5.5, respectively. As expected it increases with the  molecular weight. A comparison to the dPGS size estimated from the sulfate peak in Fig.~\ref{rdf_dpgs}(a) from the simulation, $2r_d$, shows not a good agreement, probably because the correlated solvation layer, i.e., the Stern layer, is relatively thick and reaches 
out further into the bulk.  The analysis in the next section, where an effective dPGS radius based on the ionic charge distribution is calculated, 
fully supports this conjecture.

\subsection{Electrostatic properties of dPGS}

\begin{figure}[h]
\includegraphics[scale=0.64]{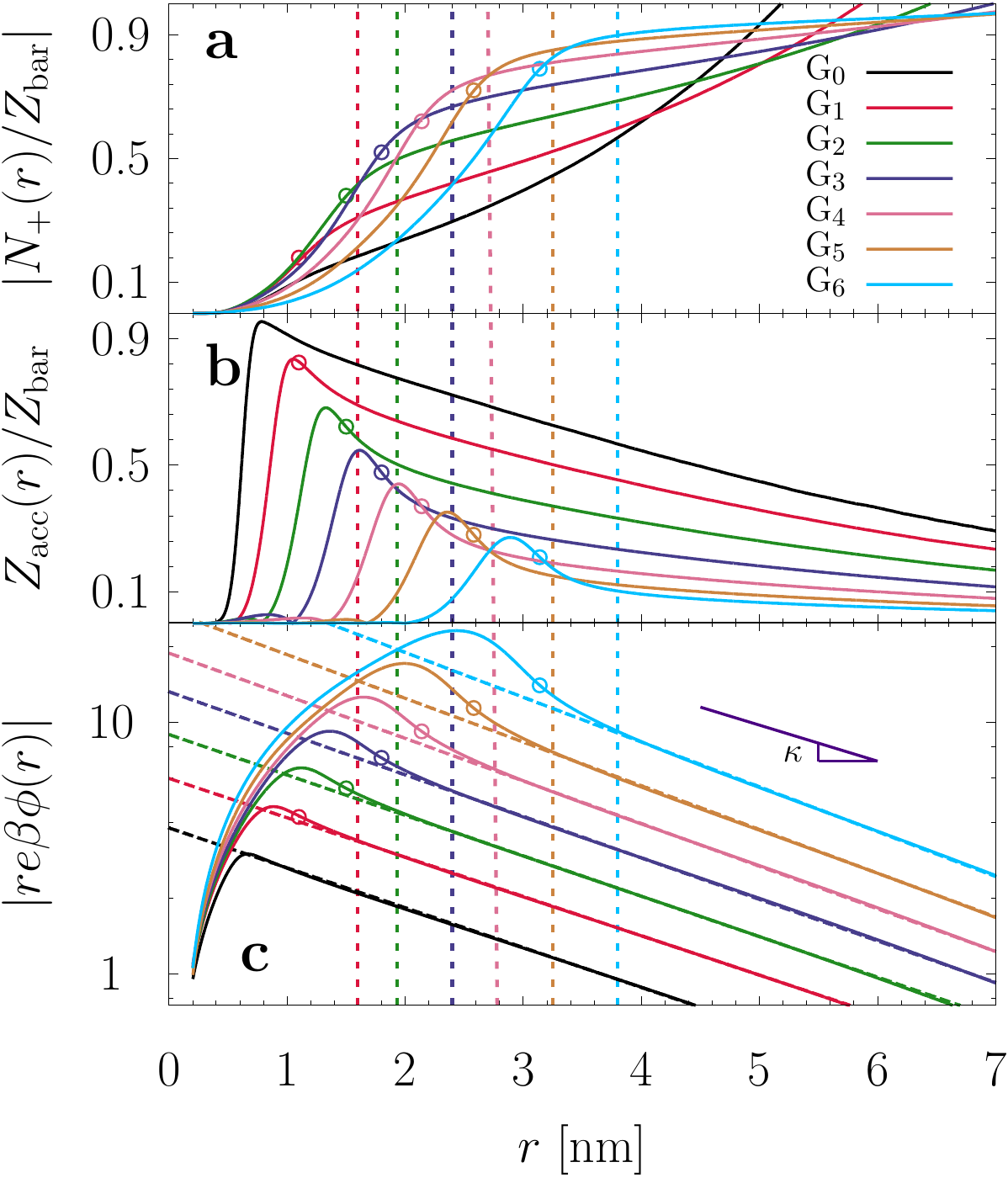}
\caption{ 
Distance--resolved (a) normalized running coordination number of counterions $N_{+}$/$Z_{\rm bar}$, 
(b) normalized accumulated effective charge $Z_{acc}(r)/Z_{\rm bar}$, 
(c) the product of distance $r$ and dimensionless radial electrostatic potential $|e\beta\phi|$.
We put a log-scale in (c) at the $y$-axis to examine the exponential decay. 
The long dashed line in (c) are fits according to the DH potential $\phi_{\rm DH}$, see Eq.(\ref{DH_solution}).
The vertical dashed lines signify the effective radius $r_{\rm eff}$ of dPGS,
whereas the circles mark the inflection radius $r_{\rm inf}$.
The dPGS generation for that plot ranges from $0$ to $6$ in condition of salt concentration $c_0=10$ mM. 
The exponential decay with standard inverse DH length $\kappa = 0.33$~nm$^{-1}$ derived by DH theory is plot in panel (c) as a further comparison.
}
\label{phi_dpgs}
\end{figure}

\subsubsection{Electrostatic potential, charge renormalization and effective `charge' radius}

From the ionic density profiles we can now in detail investigate local charge accumulation and electrostatic potential distributions. 
The accumulated number $N_{+} (r)$ of counterions and the locally total accumulated dPGS  charge $Z_{\rm acc} (r)$ (i.e., the local effective charge according to Eq.~(\ref{acc})) are presented in Fig~\ref{phi_dpgs}(a) and Fig~\ref{phi_dpgs}(b), respectively. It is found that both $N_{+} (r)$ and $Z_{\rm acc} (r)$ increase sharply with distance $r$ from the dPGS core due to the rising accumulation of charged beads. While $N_{+} (r)$ naturally rises, $Z_{\rm acc}(r)$ jumps to a maximum at a distance $r_Z$ and drops gradually. At this distance, a large portion of the sulfate charges are neutralized by counterions. As implied by Fig~\ref{phi_dpgs}(b), we find for instance that more than 70$\%$ of the charges at  $r = r_Z$ for G$_6$ are compensated by bound counterions.  This charge renormalization effect has been extensively studied at hand of simple charged spheres
with smooth surfaces, and theories for the effective charge and size have been developed~\cite{Belloni1998, DavidA.J.Gillespie2014,Ohshima1982, Bocquet2002, Manning2007}. One important outcome is that one can define the critical location for counterion-condensation in terms of the inflection point in a plot of $Z_{\rm acc}$ as a function of the inverse radial 
distance $1/r$~\cite{Belloni1998, DavidA.J.Gillespie2014}.  The equation ${d^2 Z_{\rm acc}}/{d(1/r)^2}|_{r=r_{\rm inf}} = 0$ leads to the inflection radius, $r_{\rm inf}$, within ions are deemed condensed. We marked $r_{\rm inf}$ by circle symbols for all generations in Fig~\ref{phi_dpgs}. 
Note that $r_{\rm inf}$ is larger than $r_Z$ and could be used to read off an effective charge size and charge of the macromolecule. 

A more practical concept to define an effective size and charge is to quantitatively map the double-layer decay of the potential onto the basic DH-theory, Eq.~(\ref{DH_solution}).  Fig.~\ref{phi_dpgs}(c) plots the rescaled potential $|re\beta\phi|$ versus distance $r$ in a log--linear scale.
For the potential far away a homogeneously charged sphere, the DH potential $\phi_{\rm DH}$ with a renormalized charge should apply,  yielding 
an exponential decay ${e^{-\kappa(n) r}}/{r}$ attributed to the electrostatic screening.
The plot indeed shows the expected linear decay at large separations unambiguously for all presented dPGS with a slope as expected 
to be close to the standard inverse DH length $\kappa = 0.33$~nm$^{-1}$ for the salt concentration $c_0=10$~mM. In detail, we find slopes of 
$\kappa(n)$ to monotonically increase with $n$ from $\kappa(0) = 0.36$~nm$^{-1}$  to $\kappa(6) = 0.41$~nm$^{-1}$. That slight 
increase is due to the increasing number of counterions in the finite system which also contribute to screening. 
In contrast to the simple exponential decay at large separations, for smaller distances the potential term $|re\beta \phi(r)|$ climbs up quickly with decreasing distance $r$ to a maximum before it decays to almost vanishing potential close to the dPGS core.  This highly nonlinear behavior is expected from the high electrostatic and steric correlations between sulfate beads and counterions in this dense Stern layer. 

As indicated by Fig.~\ref{phi_dpgs}(c), the potential can now be naturally divided into two parts:
a DH-regime $r > r_{\rm eff}$, where the DH potential describes correctly the potential, and a non-DH 
regime $r_d < r < r_{\rm eff}$, where a non-monotonic and highly non-exponential behavior is revealed.
In that sense, $r_{\rm eff}$ now acts as a measure of the dPGS effective radius at which we 
can attain an effective dPGS charge $Z_{\rm eff}$~\cite{Belloni1998} (see also the SI).  We depict the position of the 
DH radius $r_{\rm eff}$ by vertical dashed lines in Fig~\ref{phi_dpgs}.
We list $r_{\rm eff}$ and $Z_{\rm eff}$ in the Table~\ref{dPGS_para} for generations 0 to 6. 
We find an increase of both $r_{\rm eff}$ and $Z_{\rm eff}$ with dPGS generation index and substantial charge renormalization effects. 
For instance, the bare charge for G$_6$--dPGS is $-384$~$e$ (still presents at a radius $r_d$, cf. Fig. 5(b)) 
is effectively renormalized to $Z_{\rm eff} = -32.9$~$e$ at large distances $r>r_{\rm eff}$. 
An exception is G$_0$ in which case we find $r_d \approx r_{\rm eff}$ and hardly any renormalization by condensed counterions takes places. 
In agreement, the accumulated counterions profile $N_{+}$ for G$_0$ in Fig~\ref{phi_dpgs}(a) reveals DH behavior almost in the full 
range of $r$ and no inflection radius could be identified. 

Note that both $r_{\rm inf}$ and $r_{\rm eff}$ can in principle be taken as definition for the effective size and charge 
of the charge-renormalized sphere. Although the difference between them appears not so large in Fig.~\ref{phi_dpgs},
still a significant charge renormalization happens in between as the gradient $d N_{+}(r)/d r$ at $r_{\rm inf}$ is relatively large, \ie, 
there is a marked density decrease of the counterions from distance $r_{\rm inf}$ to $r_{\rm eff}$. 
In the following, we base our analysis only on $r_{\rm eff}$ as we believe that the inflection point criterion
holds only for more idealized systems (smooth surfaces, no salt). The procedure to obtain $r_{\rm eff}$ rests 
on the assumption that we can  treat the dPGS as simple DH spheres, so it exactly serves our purpose.  
Interestingly, a comparison of the corresponding effective diameter, $2r_{\rm eff}$ to the size measured in the Zetasizer 
experiments in Fig~\ref{DLS}(a) shows satisfying agreement. This demonstrates that the thickness of the correlated Stern layer 
in our simulation is of reasonable size and resembles the size of the bound solvation layer revealed by the experiments.

\begin{figure}[h]
\includegraphics[scale=0.66]{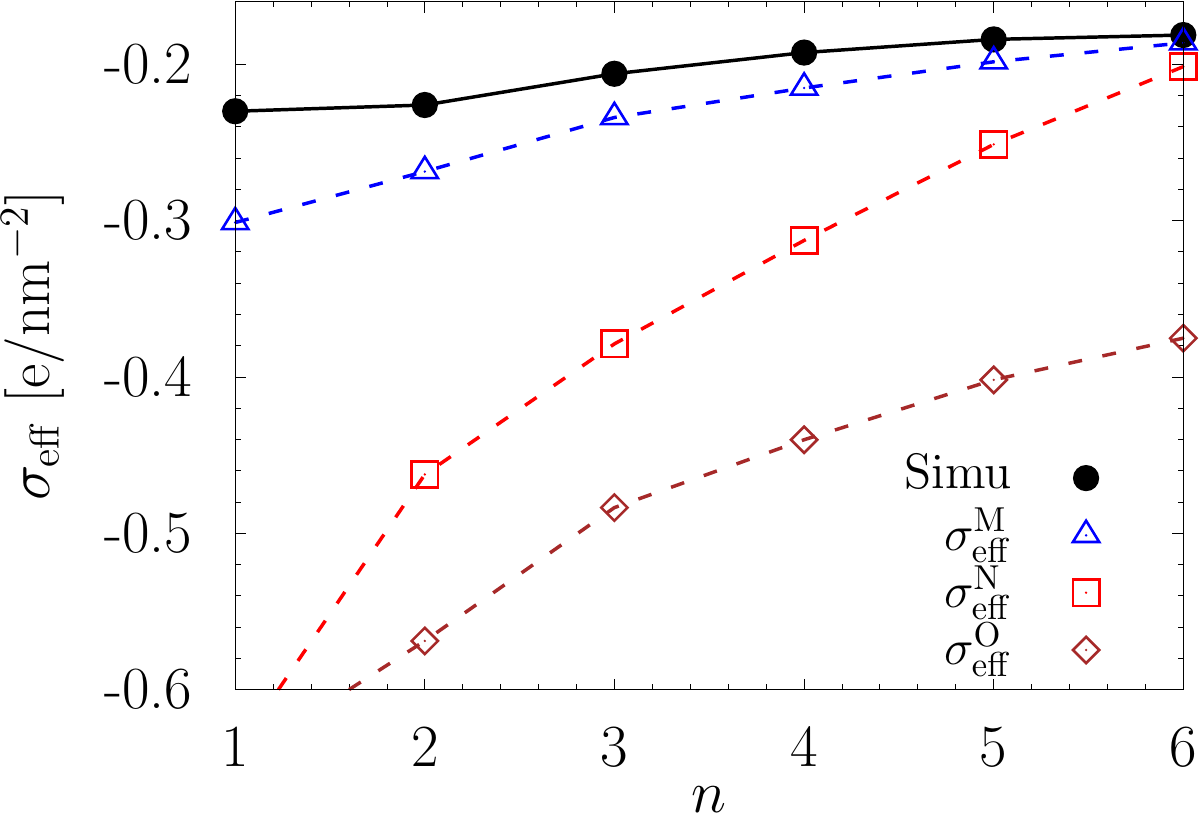}
\caption{dPGS effective surface charge density $\sigma_{\rm eff}= Z_{\rm eff}/(4\pi r^2_{\rm eff})$ versus dPGS generation number~$n$.
The solid circles are simulation results ranging from G$_1$ to G$_6$. The empty upward triangles, squares,
diamonds are effective charge based on various theories as expressed by $\sigma_{\rm M}, \sigma_{\rm N}, \sigma_{\rm O}$ 
in eqs.~(\ref{Manning}), Eq.~(\ref{Netz}), and Eq.~(\ref{Oshima}), respectively.
}
\label{sigma}
\end{figure}

\subsubsection{Effective surface charge density and potential} 

Considering the intrinsic radius $r_d$ and the DH radius $r_{\rm eff}$, the estimates of $Z_{\rm bar}$ and $Z_{\rm eff}$ lead to the dPGS bare 
surface charge density $\sigma_d$ and effective surface charge density $\sigma_{\rm eff}= Z_{\rm eff}/(4\pi r^2_{\rm eff})$, respectively.
Numbers are also summarized in Table~\ref{dPGS_para}. We find a monotonic decrease on $\sigma_d$ with generation $n$, in response to the growing number of the surface beads.  (The small portion of backfolding terminal beads increasing with generation decreases the surface charge valency slightly but not significant).  Due to the large counterion-binding and charge renormalization effect, the effective charge density $\sigma_{\rm eff}$ is about one order of magnitude smaller than the bare one. Interestingly, it virtually remains constant, even slightly decreases from $-0.23$~$e$ nm$^{-2}$ for G$_1$ to a saturated value $-0.18$~$e$ nm$^{-2}$ for G$_5$ and G$_6$.    Experiments of carboxyl-terminated dendrimers at $pH$ much larger than the $pKa$ (i.e., almost full ionization) also found higher effective charge densities of a lower generation G$_2$ than for G$_5$~\cite{Huang2000}.  

The results for the effective surface charge can be compared to available theories of charge renormalization of highly charged spheres, typically valid in low or high salt limits.  Early approaches are based on approximate solutions of the nonlinear Poisson-Boltzmann equation for isolated spheres at infinite dilution, e.g., improvements of the now classical Ohshima potential~\cite{Ohshima1982} lead to~\cite{Bocquet2002}
\begin{equation}
\sigma^{\rm O}_{\rm eff} = \frac{-2e[1+\kappa(n) r_{\rm eff}]^2}{\pi \lB r_{\rm eff} [1+2\kappa(n) r_{\rm eff}]}
\label{Oshima}
\end{equation}
and should be valid for large $\kappa r_{\rm eff}\gtrsim 1$, i.e., large spheres and/or high screening by salt. 
In the framework of standard counterion-condensation theory, Manning later derived a different but related 
expression for the  saturation surface charge density of an isolated sphere in the same regime 
($\kappa r_{\rm eff}\rightarrow 0$) as~\cite{Manning2007}
\begin{equation}
\sigma^{\rm M}_{\rm eff} = \frac{e[1+\kappa(n) r_{\rm eff}]\ln[\kappa(n) \lB]}{2\pi \lB r_{\rm eff}}.
\label{Manning}
\end{equation}
In the other limit ($\kappa r_{\rm eff}\rightarrow 0$), Netz {\it et al.}~\cite{Netz2003} instead provide an estimate on the effective charge density explicitly dependent on the bare charge valency, $Z_{\rm bar}$,  by means of variational techniques, via
\begin{equation}
\sigma^{\rm N}_{\rm eff} = \frac{-e}{4\pi r_{d} \lB} \ln{\left( \frac{\lB |Z_{\rm bar}|}{r_{d}^3 \kappa^2} \right)}.
\label{Netz}
\end{equation}
For our dPGS systems, intermediate values $0.25 < \kappa r_{\rm eff} < 1.6$ are established, for which an accurate analytical description
apparently is difficult to achieve. We plot $\sigma_{\rm eff}$ in terms of generation number $n$, together with $\sigma^{\rm M}_{\rm eff}, \sigma^{\rm N}_{\rm eff}$ and $\sigma^{\rm O}_{\rm eff}$ in Fig.~\ref{sigma}. We find that simulation and theory both yield the same trend, i.e., the absolute effective surface charge density decreases with generation $n$.  Given the enormous charge renormalization effects of about one order of magnitude, the agreement to all theories is actually satisfactory, especially for the Manning theory.  The relative error with respect to the magnitude of the bare surface charge is thus less than 10\%. Based on this empirical comparison, the Manning approach can thus serve as a simple and analytical extrapolation to other systems and experiments.

Correspondingly, we define the  dPGS surface potential $\phi(r_{\rm eff})$, also summarized in Table~\ref{dPGS_para}. 
Similarly as the effective surface charge the surface potential is staying relatively constant with generation number. 
An inspection and comparison to $\zeta$-potentials derived from our electrophoretic mobility experiments is made in Fig~\ref{DLS}(b).
Recall that the shear plane, where the $\zeta$-potential is located, should lie beyond the Stern layer which 
refers to a position very close to $r_{\rm eff}$ where our effective surface potential $\phi(r_{\rm eff})$ is in the simulation~\cite{Lyklem1995, Ahualli2016}. Indeed, as we see in  Fig~\ref{DLS}(b), $\phi(r_{\rm eff})$ reproduces the experimental $\zeta$-potential very well at all generations. 
The sources of the remaining deviations can be of various origin, e.g., missing explicit water contributions to the electrostatic potential in the CG simulations or simply the lack of the exact knowledge of the shear plane.  If, for instance, we assume an up- or down-shift of the location of the shear plane only about one solvation layer,  say to be $\simeq r_{\rm eff} \pm 0.3$~nm, the experimental range is well 
matched, cf. Fig~\ref{DLS}(b). Note also that in the experiments not directly the potential is measured but the
 electrophoretic mobility (presented in the SI) and their conversion is based on idealized models~\cite{Hunter, Huang2000}.

\begin{figure}[h]
\includegraphics[scale=0.67]{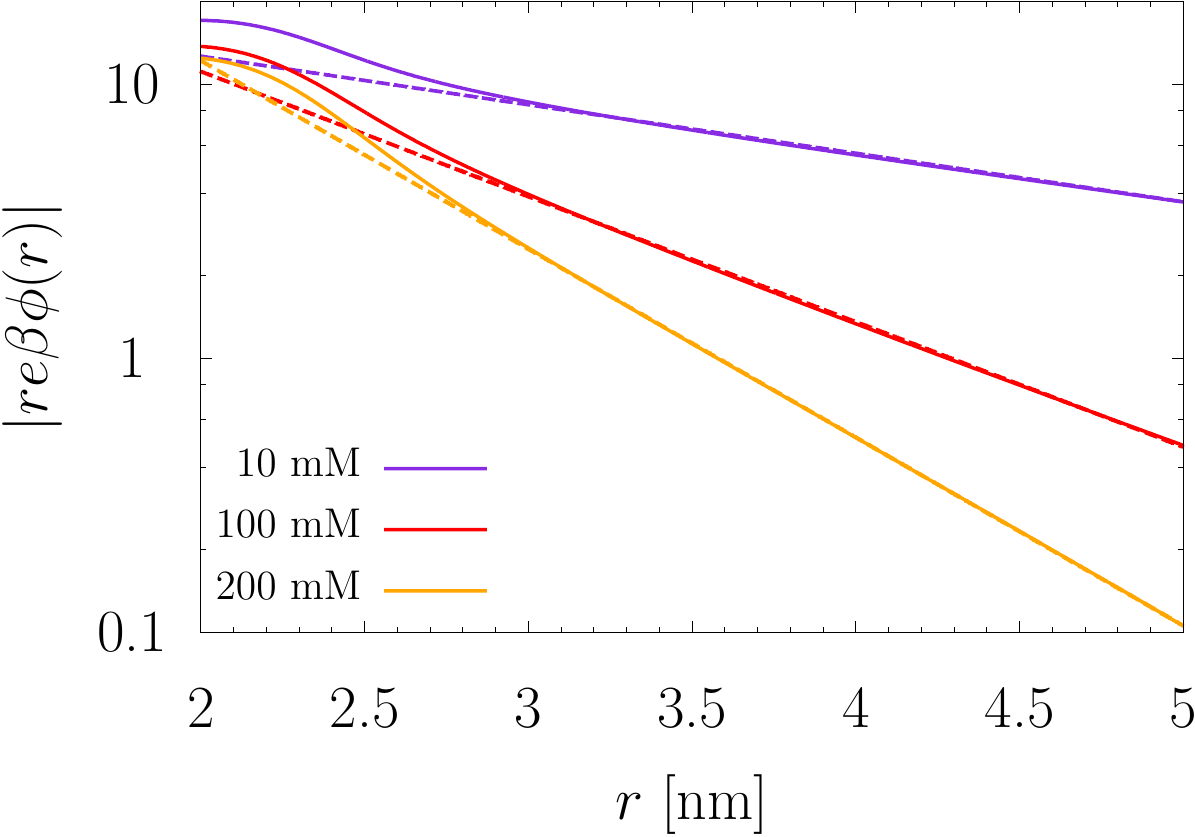}
\caption{
Log-linear plot of the distance-resolved G$_5$-dPGS radial electrostatic potential times distance, $|re\beta\phi|$, as in Fig. \ref{phi_dpgs}(c), but now for different 
salt concentrations $c = 10$, 100, and 200 mM.  The dashed lines depict the corresponding DH potential $\phi_{\rm DH}$ after fitting to the linear decay regime. 
}
\label{phi_c}
\end{figure}

We finally inspect the dependence of the effective dPGS radius $r_{\rm eff}$ and charge density $\sigma_{\rm eff}$ on the salt concentration. 
We plot the radial electrostatic potential $|re\beta\phi|$ for G$_5$-dPGS under three salt concentrations $c = 10$ mM, 100 mM, and 200 mM 
in Fig.~\ref{phi_c}.  As one can see the exponential regime $\exp(-\kappa r)$ for large distances survives for all cases.
Interestingly, a similar $r_{\rm eff} \simeq 3.2$~nm is found with hardly any dependence on the salt concentration.
The effective surface charge density goes down from $\sigma_{\rm eff} = -0.18~e$/nm$^{-2}$ (10 mM) to $\sigma_{\rm eff} = -0.14~e$/nm$^{-2}$ (100 mM) and $\sigma_{\rm eff} = -0.10~e$/nm$^{-2}$ at 200 mM.  Eq.~(\ref{Oshima}) predicts an opposite trend than that and than Eqs.~(\ref{Manning}) and~(\ref{Netz}). The latter two treatments consistently point to the generic screening effect 
leading to a smaller surface potential $\phi(r_{\rm eff})$ for higher salt concentrations. 

\section{Conclusions}
In summary, we have investigated the electrostatic (surface) properties of highly charged dendrimers for various generations at hand of the biomedically important dPGS polyelectrolyte.  We have defined an effective charge, effective surface charge and potential of dPGS for various generations and salt concentrations by a direct mapping procedure of the calculated electrostatic potentials onto the long-range Debye-H{\"u}ckel-like electrostatic decay which defines the effective charge in its most practical level.  The dPGS effective radius $r_{\rm eff}$ is accordingly addressed as a distance separating double-layer and condensation regimes and therefore gives the dPGS effective charge without ambiguity.  Evidently, with that procedure the effective charge and the surface potential and their trends with generation can be consistently described by counterion-condensation theory and show very good agreement with new experimental $\zeta$-potential  measurements as well.

In future, our model can be easily applied to dPGS-involved intermolecular interaction studies in biological context (e.g., binding to proteins or membranes), quantifying electrostatic interactions, in particular counterion-release effects on binding~\cite{Yu2015, Yigit2015b,Yigit2016}.  Those studies could serve as important references to guide experiments and optimize dPG-based particles as a potent anti-inflammatory drug in
 biomedical applications. In particular, the strength of specific counterion binding and condensation has significant influence on binding affinity of dPGS or other charge-functionalized polyglycerol-based dendrimers, such a carboxylated or phosphorylated dPG~\cite{Weinhart2011, Weinhart2011a}.
 \\

\begin{acknowledgement}
Xiao Xu thanks the Chinese Scholar Council for financial support.
The authors acknowledge fruitful discussions with Rohit Nikam and Rafael Roa Chamorro. 
\end{acknowledgement}

\setlength{\bibsep}{0pt}
\providecommand{\latin}[1]{#1}
\providecommand*\mcitethebibliography{\thebibliography}
\csname @ifundefined\endcsname{endmcitethebibliography}
  {\let\endmcitethebibliography\endthebibliography}{}

\end{document}